\documentclass[a4paper,11pt]{article}

\pdfoutput=1
\usepackage{booktabs}

\usepackage{lineno}
\usepackage[colorlinks=false]{hyperref}
\usepackage{bm}
\usepackage{graphicx}
\usepackage{amssymb}
\usepackage{amsmath}
\usepackage{tabularx}
\usepackage{natbib}
\usepackage{floatrow}
\usepackage{caption}
\usepackage{amsmath}
\usepackage{amsfonts}
\usepackage{amssymb}

\newcommand*{\abs}[1]{\left|{#1}\right|}
\newcommand*{\bracket}[1]{\left({#1}\right) } 
\newcommand*{\sbracket}[1]{\left[ {#1}\right] } 

\modulolinenumbers[5]

\usepackage{authblk}

\usepackage[left=15mm,right=15mm,top=1.5cm,bottom=1.5cm,includeheadfoot]{geometry}
\usepackage{multirow}
\usepackage{siunitx}
\usepackage[labelformat=simple]{subcaption}

\newcommand*{\tensor}[1]{\bar{\bar{#1}}}

\newcommand{\beginsupplement}{%
	\setcounter{table}{0}
	\renewcommand{\thetable}{S\arabic{table}}%
	\setcounter{figure}{0}
	\renewcommand{\thefigure}{S\arabic{figure}}%
}

\setcitestyle{square,numbers}

\begin{document}

\title{Silicon substrate significantly alters dipole-dipole resolution in coherent microscope}


\author[1]{Zicheng Liu}

\author[1]{Krishna Agarwal}

\affil[1]{\scriptsize Department of Physics and Technology, UiT The Arctic University of Norway, NO-9037 Troms\o, Norway}

\maketitle

\abstract{
Influences of a substrate below samples in imaging performances are studied by reaching the solution to the dyadic Green's function, where the substrate is modeled as half space in the sample region. Then, theoretical and numerical analysis are performed in terms of magnification, depth of field, and resolution. Various settings including positions of dipoles, the distance of the substrate to the focal plane and dipole polarization are considered. Methods to measure the resolution of $z$-polarized dipoles are also presented since the modified Rayleigh limit cannot be applied directly. The silicon substrate and the glass substrate are studied with a water immersion objective lens. The high contrast between silicon and water leads to significant disturbances on imaging.}

\section{Introduction}
Fluorescence imaging \cite{stockert2017fluorescence} suffers from photobleaching and blinking, and the imaging duration is limited by the photochemically toxic environment. Therefore, label-free imaging \cite{Marx2019It} is desired but requires improvements in resolution. With no or little modifications of the existing instruments, computational techniques \cite{agarwal2016multiple,crocco2012linear,abubakar2002imaging} are used to enhance the resolution by making use of the point spread properties, which are commonly studied through the point spread function and aberration functions in optics. 

\begin{figure}[!ht]
	\centering
	\begin{subfigure}{.4\textwidth}
		\centering
		\includegraphics[height=0.5\linewidth]{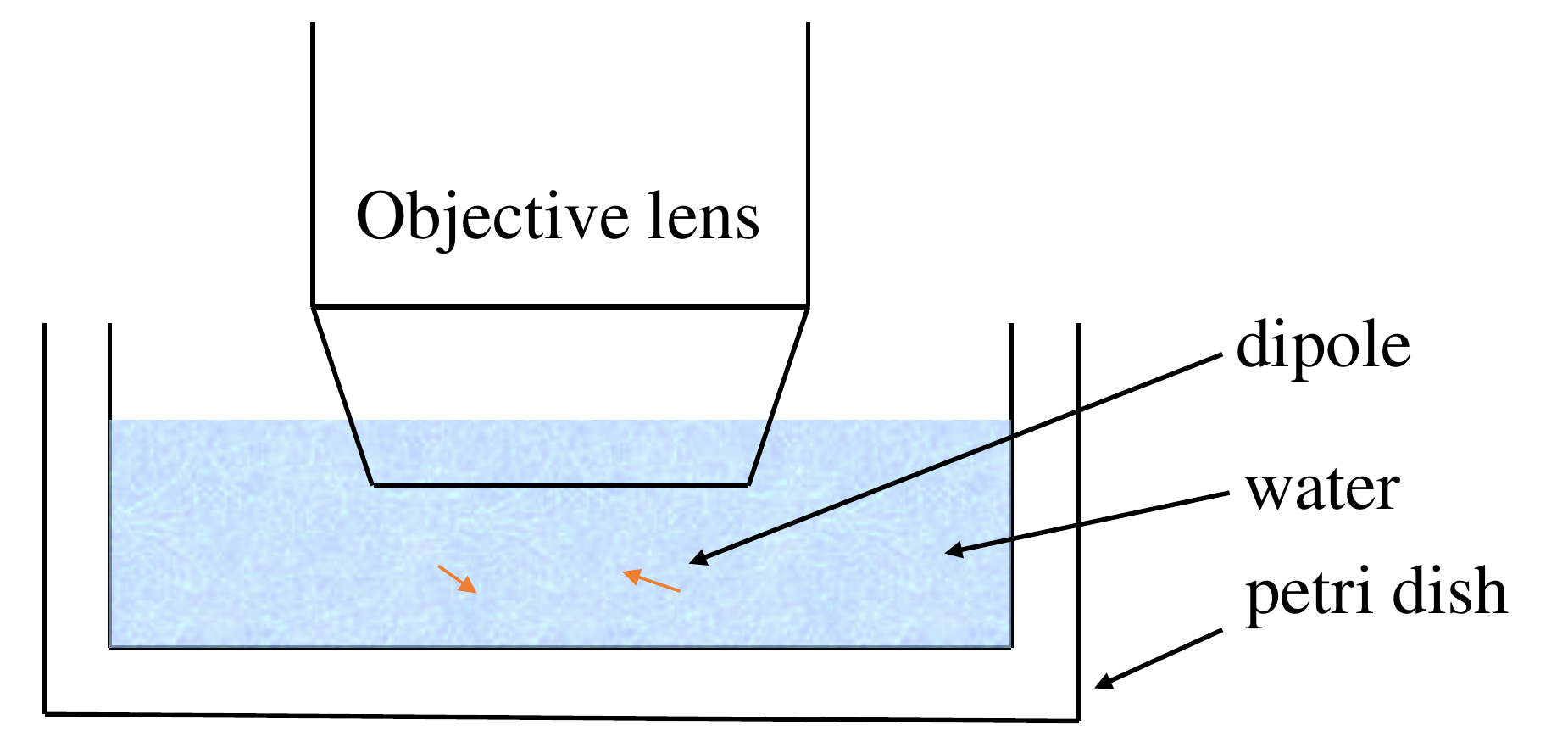}
		\caption{}
		\label{fig:microscope_petri_dish}
	\end{subfigure}
	\hspace{0.05\linewidth}
	\begin{subfigure}{.4\textwidth}
		\centering
		\includegraphics[height=0.5\linewidth]{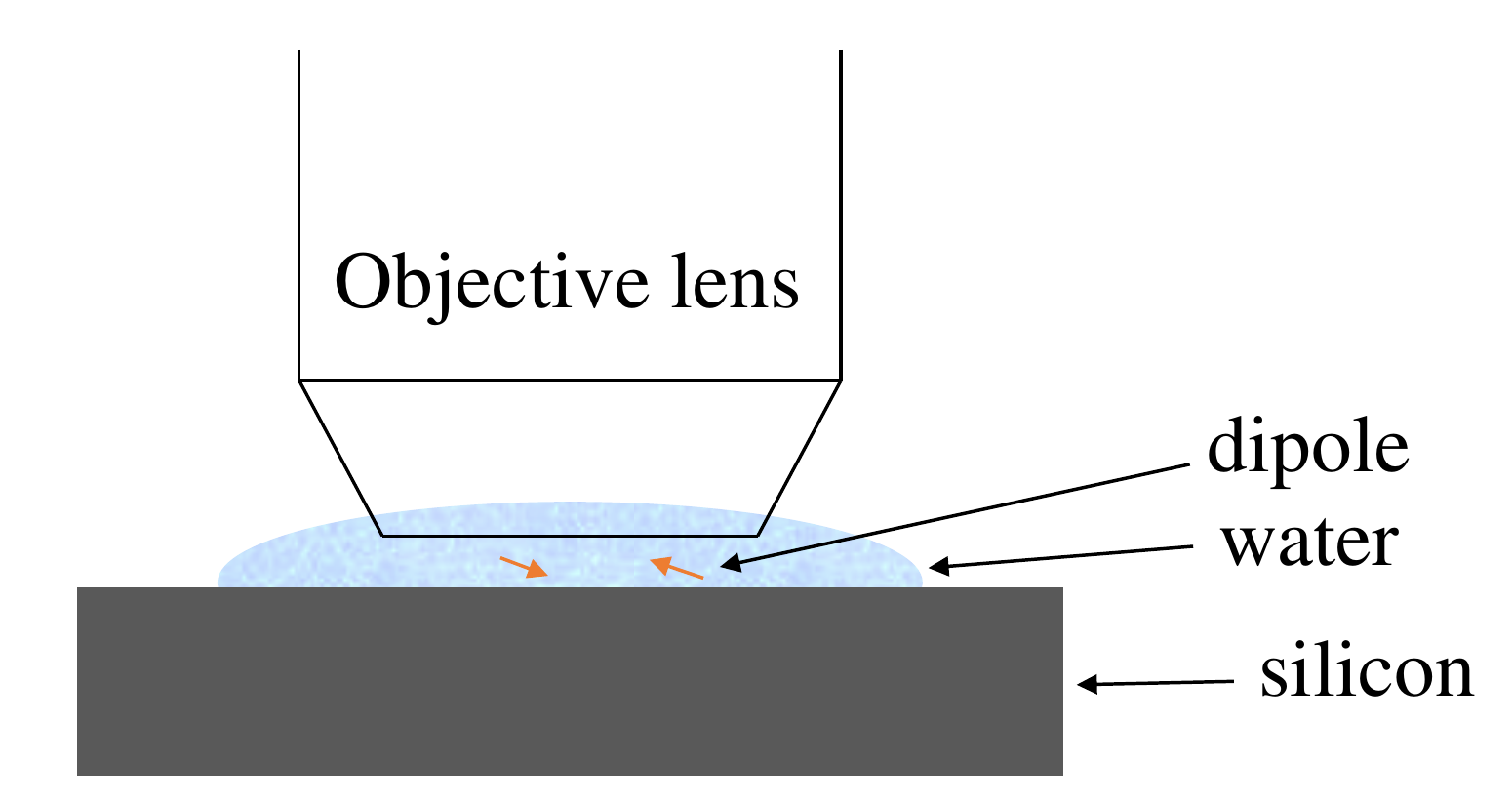}
		\caption{}
		\label{fig:microscope_slicon}
	\end{subfigure}
	\caption{Sketch of microscope with a water immersion objective lens when samples are (a) in petri dish and (b) on the silicon substrate, respectively.}
	\label{fig:microscope}
\end{figure}

A full electromagnetic wave approach, rather than ray optics (geometric optics), increases the applicability of the computational models of microscope to high NA systems. This was for example demonstrated before in the case of solid immersion lens \cite{hu2011dyadic,chen2012resolution,chen2013complete}, where computational modeling used for identification of the suitable pinhole dimension that allows better resolution by balancing the collection of light from longitudinal and lateral dipoles induced in the sample region. A more general version, without assuming solid immersion lens is available in \cite{novotny2012principles}. While general, it lacks one inevitable aspect of microscopy, especially when used for biological imaging. This aspect is the presence of an interface due to the resting surface of sample, for example petri dish in Fig.~\ref{fig:microscope_petri_dish} and the silicon substrate in Fig.~\ref{fig:microscope_slicon}  in modern photonics-based microscopes \cite{archetti2019waveguide,ahmad2020sub}. The importance of modeling the interface has been recognized before, and incoherent point spread functions of optical microscopes have been derived \cite{gibson1991experimental}. However, we are not aware of analogous 3D full-wave dyadic green function (DGF) for a coherent microscopy system. We do note that there have been works related to layered medium \cite{guo2007multilayered} in optical systems for optical memory and lithography. However, for a microscopy system involving the conventional microscope-objective and tube lens pair is currently not simulated. Here, we present the DGF of half space emulating an air or liquid-dipping objective based coherent microscopy system where the half space corresponds to the glass surface or silicon surface. 

Expanding the field emitted by a dipole into plane waves, the DGF of the concerned microscopy system is solved
by analyzing the reflection and refraction behaviors of each plane-wave component and integrating all rays reaching the image region. Based on the closed-form solution of DGF, the lateral and longitudinal resolutions are studied with various settings. Quantifying the resolvability by saddle-to-peak ratio, the modified Rayleigh limit is used as the resolution criterion. Further, since $z$-polarized dipoles are shown as annuli \cite{novotny2012principles} and the computation of saddle-to-peak ratio is not straightforward, efforts are made here to apply the modified Rayleigh limit to $z$-polarized dipoles. The influences of the substrate on the depth of field are also discussed. 

\section{Setup and notations}
\begin{figure}[!h]
	\centering
	\includegraphics[width=0.7\linewidth]{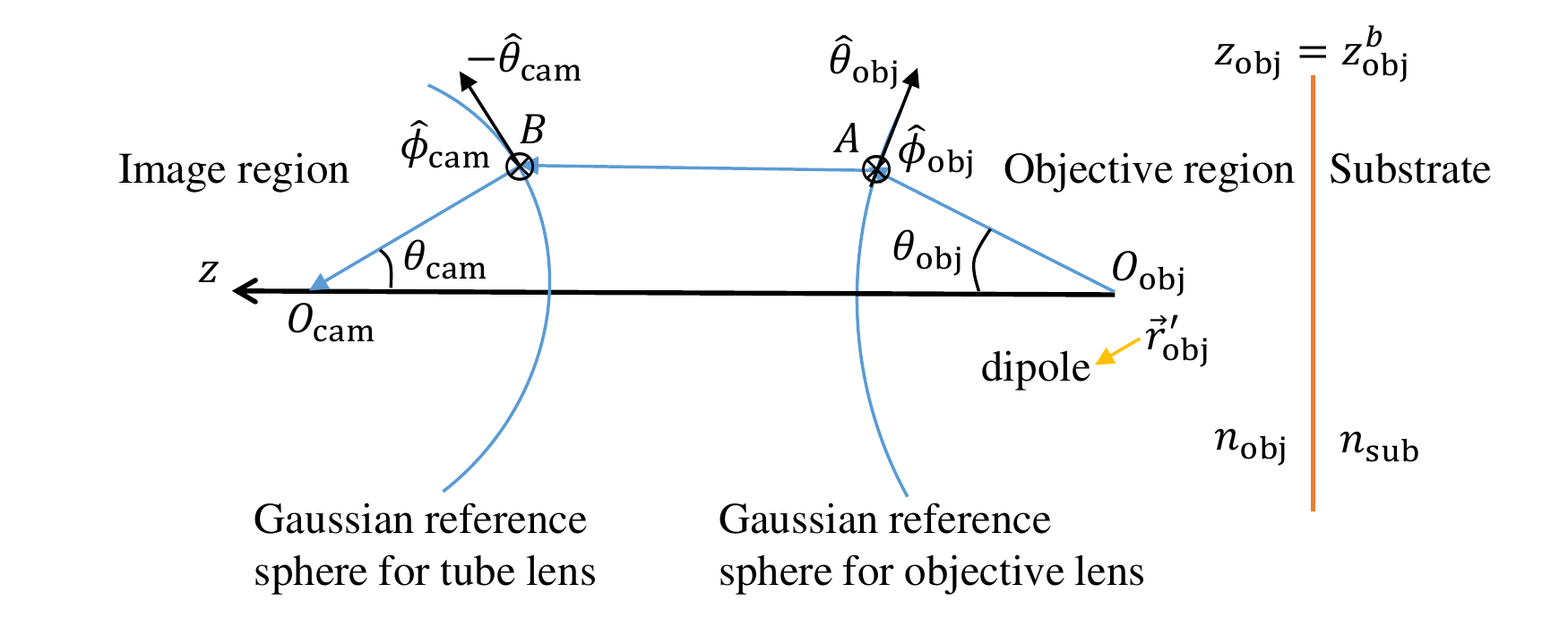}
	\caption{Sketch of wave propagation in the aplanatic system with a half space.}
	\label{fig:sketchMicroscope}
\end{figure}
Fig.~\ref{fig:sketchMicroscope} presents the schematic diagram and the important notations of the concerned microscopy system. Taking the optical axis as the $z$ axis, two local coordinate systems are built {\color{black}whose origins are located} at the focal point{\color{black}s} of the objective region and {\color{black}the} image region, respectively. {\color{black}These regions are respectively created by the objective lens (focused on the sample) and the tube lens (focused on the camera).} The notation of quantities in the objective region {\color{black}employs} the subscript ``obj", while quantities in the image region{\color{black}, where camera sensors (scientific CMOS, CCD or emCCD) are positioned,} are subscripted by ``{\color{black}cam}". Here, the general derivation is being made for the region close to the image plane and the center of camera pixel is used to represent the pixel assuming a pixel to be point-like detector. However, generalization to a large pixel or optical detector size (such as photodetector or multiphoton diode) can be made by integrating the computed intensity over the area of the detection element.

In the objective region, $O_\mathrm{obj}$ denotes the focal point, $n_\mathrm{obj}$ the refractive index of the medium in which sample is kept, $k_\mathrm{obj}$ the wavenumber, $f_\mathrm{obj}$ the focal length. $\vec{r}^\prime_{\mathrm{obj}}$ denotes the position of a dipole with Cartesian coordinates $[x^\prime_\mathrm{obj},y^\prime_\mathrm{obj},z^\prime_\mathrm{obj}]$. $\vec{r}_\mathrm{obj}^o$ denotes an example observation point in the objective lens region with Cartesian coordinates $[x_\mathrm{obj}^o,y_\mathrm{obj}^o,z_\mathrm{obj}^o]$ and spherical coordinates $[r_\mathrm{obj}^o,\theta_\mathrm{obj}^o,\phi_\mathrm{obj}^o]$. This observation point in the objective region is defined to facilitate the derivation of the DGF and is not used in the final formulation of DGF. Remark that the vector quantities are noted by putting an arrow above and $r_\mathrm{obj}$ stands for the radial distance to $O_\mathrm{obj}$. The notations in the image region are similarly defined. The {\color{black}substrate is modeled as the lower half space and} characterized by the position of the substrate interface $z_\mathrm{obj}=z_\mathrm{obj}^b$ (i.e., parallel with the $x$,$y$ plane) and the refractive index $n_\mathrm{sub}$. The angular semiaperture of the objective lens is $\theta_\mathrm{obj}^\mathrm{max}$. Both the objective lens and the tube lens are represented by Gaussian reference spheres (GRS) \cite{novotny2012principles}.

\section{Solution to dyadic Green's function}
\label{sec:DGF_microscope}

As sketched in Fig.~\ref{fig:sketchMicroscope}, a unit dipole polarized along an arbitrary direction $\hat{a}$ locates in the objective region. After reflections by the substrate interface, the emitted waves propagate through the objective lens and the tube lens, and finally recorded {\color{black}at pixel locations on camera array}. We denote the field solution as $\vec{G}(\hat{a})$. The dyadic Green's function (DGF) is defined by $\tensor{G}=[\vec{G}(\hat{x}),\vec{G}(\hat{y}),\vec{G}(\hat{z})]$, which is a $3\times 3$ tensor. Following the derivation in \cite{GreenKong}, the solution to the DGF of the half space is a superposition of TE (s-polarized) and TM (p-polarized) parts,
\begin{equation}
\label{eq:fieldObject_sphercialCoord}
\tensor{G}(\vec{r}_{\text{obj}}^o,\vec{r}^\prime_{\text{obj}}) = \dfrac{1}{8i\pi^2}\int\int_{-\infty}^{+\infty}dk_xdk_y\dfrac{1}{k^{z}_{\text{obj}}}\sbracket{\hat{\phi}_\mathrm{obj}\bracket{\vec{k}^+_\mathrm{obj}}\vec{K}^{\text{TE}}+\hat{\theta}_\mathrm{obj}\bracket{\vec{k}^+_\mathrm{obj}}\vec{K}^{\text{TM}}}e^{i\vec{k}^+_{\text{obj}}\vec{r}_{\text{obj}}^o},
\end{equation}
in which 
\begin{subequations}
	\begin{equation}
	\label{eq:KTE}
	\vec{K}^{\text{TE}} = -\hat{\phi}_\mathrm{obj}\bracket{\vec{k}^+_\mathrm{obj}}e^{-i\vec{k}^+_\mathrm{obj}\vec{r}^\prime_{\text{obj}}}-\hat{\phi}_\mathrm{obj}\bracket{\vec{k}^+_\mathrm{obj}}R^{\text{TE}}e^{-i\vec{k}^-_{\text{obj}}\vec{r}^\prime_{\text{obj}}},
	\end{equation}
	\begin{equation}
	\label{eq:KTM}
	\vec{K}^{\text{TM}} = -\hat{\theta}_\mathrm{obj}\bracket{\vec{k}^+_\mathrm{obj}}e^{-i\vec{k}_{\text{obj}}^+\vec{r}^\prime_{\text{obj}}}-\hat{\theta}_\mathrm{obj}\bracket{\vec{k}^-_\mathrm{obj}}R^{\text{TM}}e^{-i\vec{k}_{\text{obj}}^-\vec{r}^\prime_{\text{obj}}}.
	\end{equation}
\end{subequations}
The first terms of Eq.~\eqref{eq:KTE} and \eqref{eq:KTM} stand for the wave emitted by the dipole and the second terms for the wave reflected by the substrate interface. $\vec{k}_{\text{obj}}^\pm = k_x\hat{x}+k_y\hat{y}\pm k^z_{\text{obj}}\hat{z}$ denote the wave vector of the integrated plane-wave components. $k^z_{\text{obj}} = \sqrt{k_{\text{obj}}^2-k_x^2-k_y^2}$ has a non-negative imaginary part when $k_x^2+k_y^2>k_{\text{obj}}^2$. $\hat{\phi}_\mathrm{obj}\bracket{\vec{k}^\pm_\mathrm{obj}}$ and $\hat{\theta}_\mathrm{obj}\bracket{\vec{k}^\pm_\mathrm{obj}}$ are defined as the unit vector of the azimuthal and elevation axis for $\vec{k}_\mathrm{obj}^\pm$. The reflection coefficients are obtained based on the continuity of tangential electric fields across the substrate interface and expressed as
\begin{subequations}
	\label{eq:solPWECoeff}
	\begin{equation}
	R^{\text{TE}}=e^{-2ik^z_{\text{obj}}z_{\text{obj}}^b}\bracket{k^z_{\text{obj}}-k^z_{\text{sub}}}/\bracket{k^z_{\text{obj}}+k^z_{\text{sub}}},
	\end{equation}
	\begin{equation}
	R^{\text{TM}} = e^{-2ik^z_{\text{obj}}z_{\text{obj}}^b}\bracket{{k^z_{\text{obj}}k_{\text{sub}}^2}-{k^z_{\text{sub}}k_{\text{obj}}^2}}/\bracket{{k^z_{\text{obj}}k_{\text{sub}}^2}+{k^z_{\text{sub}}k_{\text{obj}}^2}},
	\end{equation}
\end{subequations}
{\color{black}where $k_\mathrm{sub}$ is the wavenumber of the substrate and $k_\mathrm{sub}^z$ is similarly defined with $k_\mathrm{obj}^z$.} 

Since we consider a far-field imaging microscope, i.e., the focal length $f_{\text{obj}}\gg \lambda_{\text{obj}}$, only the far (propagating) fields  need to be considered in the integration of \eqref{eq:fieldObject_sphercialCoord}, i.e., the integral region is restricted by $k_x^2+k_y^2\le k_\mathrm{obj}^2$. With the method of stationary phase \cite{novotny2012principles}, the solution of fields (before the refraction by the objective lens) at any point on the {\color{black}objective lens} GRS surface $\vec{A}$ is solved as
\begin{equation}
\tensor{G}_\infty(\vec{A},\vec{r}^\prime_{\text{obj}})=-\dfrac{e^{ik_{\text{obj}}f_{\text{obj}}}}{4\pi f_{\text{obj}}}\sbracket{\hat{\phi}_\mathrm{obj}\bracket{\vec{A}}\vec{K}^{\text{TE}}+\hat{\theta}_\mathrm{obj}(\vec{A})\vec{K}^{\text{TM}}},
\end{equation}
Following the sine condition and the intensity law \cite{novotny2012principles}, after the refraction by the objective lens and the tube lens, the field at {\color{black}the tube lens GRS surface} $\vec{B}$ is with expression
\begin{equation}
\tensor{G}_{\infty}(\vec{B},\vec{r}^\prime_{\text{obj}})=-\dfrac{e^{ik_{\text{obj}}f_{\text{obj}}}}{4\pi f_{\text{obj}}}\sqrt{\dfrac{n_{\text{obj}}}{n_\mathrm{cam}}}\sqrt{\dfrac{\cos\theta_{\text{obj}}^A}{\cos\theta_\mathrm{cam}^B}}\sbracket{\hat{\phi}_\mathrm{cam}\bracket{\vec{B}}\vec{K}^{\text{TE}}-\hat{\theta}_\mathrm{cam}(\vec{B})\vec{K}^{\text{TM}}},
\end{equation}
where $\theta_\mathrm{obj}^A$ and $\theta_\mathrm{cam}^B$ are the elevation angle of $\vec{A}$ and $\vec{B}$, respectively. The superposition of all rays from the surface of the tube lens yields the solution of DGF, i.e.,
\begin{equation}
\tensor{G}(\vec{r}_\mathrm{cam}^o,\vec{r}^\prime_{\text{obj}}) = \dfrac{f_\mathrm{cam}e^{ik_\mathrm{cam}f_\mathrm{cam}}}{2i\pi} \int\int_{k_x^2+k_y^2\le \min\{k_{\text{obj}}^2,k_\mathrm{cam}^2\}}d\vec{k}_\bot \tensor{G}_{\infty}e^{i\vec{k}^+_\mathrm{cam}\vec{r}_\mathrm{cam}^o}\dfrac{1}{k^z_\mathrm{cam}}.
\end{equation}
The limit by the angular semiaperture is imposed by transforming the integral to the spherical coordinate system of the objective region,
\begin{equation}
\tensor{G}(\vec{r}_\mathrm{cam}^o,\vec{r}^\prime_{\text{obj}}) =\dfrac{k_\mathrm{cam}f_\mathrm{obj}^2e^{ik_\mathrm{cam}f_\mathrm{cam}}}{2i\pi f_\mathrm{cam}} \int_0^{\theta_{\text{obj}}^{\max}}\int_{0}^{2\pi} \tensor{G}_{\infty}e^{i\vec{k}^+_\mathrm{cam}\vec{r}_\mathrm{cam}^o}\dfrac{\cos\theta_{\text{obj}}}{\cos\theta_\mathrm{cam}}\sin\theta_{\text{obj}} d\phi_\mathrm{obj} d\theta_{\text{obj}}.
\end{equation}
Solving the integral about $\phi_\mathrm{obj}$ analytically, the two-dimensional integral is transformed as a one-dimensional integral about $\theta_{\text{obj}}$ only. Specifically, the expression of the DGF is then given as
\begin{equation}
\label{eq:solG}
\tensor{G}(\vec{r}_\mathrm{cam}^o,\vec{r}^\prime_{\text{obj}}) = \alpha
\begin{bmatrix}
I_{xx}^{(1)}+I_{xx}^{(2)} & I_{xy} & I_{xz}\\
I_{xy} & I_{xx}^{(1)}-I_{xx}^{(2)} & I_{yz}\\
I_{zx} & I_{zy} & I_{zz}
\end{bmatrix}
\end{equation}
where
\begin{subequations}
	\label{eq:integrals}
	\begin{equation}
	\alpha = \dfrac{k_\mathrm{cam}e^{i(k_{\text{obj}}f_{\text{obj}}+k_\mathrm{cam}f_\mathrm{cam})}}{8i\pi}\dfrac{f_{\text{obj}}}{f_\mathrm{cam}}\sqrt{\dfrac{n_{\text{obj}}}{n_\mathrm{cam}}},
	\end{equation}
	\begin{equation}
	I_{xx}^{(1)}=\int_{0}^{\theta_{\text{obj}}^{\max}}\sbracket{C_\mathrm{TE}+\cos\theta_\mathrm{cam}\cos\theta_{\text{obj}}C_\mathrm{TM}^-}J_0(\gamma)\sqrt{\dfrac{\cos\theta_{\text{obj}}}{\cos\theta_\mathrm{cam}}}\sin\theta_{\text{obj}}d\theta_{\text{obj}},
	\end{equation}
	\begin{equation}
	I_{xx}^{(2)}=\int_{0}^{\theta_{\text{obj}}^{\max}}\sbracket{C_\mathrm{TE}-\cos\theta_\mathrm{cam}\cos\theta_{\text{obj}}C_\mathrm{TM}^-}J_2(\gamma)\cos 2\psi\sqrt{\dfrac{\cos\theta_{\text{obj}}}{\cos\theta_\mathrm{cam}}}\sin\theta_{\text{obj}}d\theta_{\text{obj}},
	\end{equation}
	\begin{equation}
	I_{xy}=\int_{0}^{\theta_{\text{obj}}^{\max}}\sbracket{C_\mathrm{TE}-\cos\theta_\mathrm{cam}\cos\theta_{\text{obj}}C_\mathrm{TM}^-}J_2(\gamma)\sin 2\psi\sqrt{\dfrac{\cos\theta_{\text{obj}}}{\cos\theta_\mathrm{cam}}}\sin\theta_{\text{obj}}d\theta_{\text{obj}},
	\end{equation}
	\begin{equation}
	I_{xz}=\int_{0}^{\theta_{\text{obj}}^{\max}}-2i\cos\theta_\mathrm{cam}C_\mathrm{TM}^+J_1(\gamma)\cos\psi\sqrt{\dfrac{\cos\theta_{\text{obj}}}{\cos\theta_\mathrm{cam}}}\sin^2\theta_{\text{obj}}d\theta_{\text{obj}},
	\end{equation}
	\begin{equation}
	I_{yz}=\int_{0}^{\theta_{\text{obj}}^{\max}}-2i\cos\theta_\mathrm{cam}C_\mathrm{TM}^+J_1(\gamma)\sin\psi\sqrt{\dfrac{\cos\theta_{\text{obj}}}{\cos\theta_\mathrm{cam}}}\sin^2\theta_{\text{obj}}d\theta_{\text{obj}},
	\end{equation}
	\begin{equation}
	I_{zx}=\int_{0}^{\theta_{\text{obj}}^{\max}}i\sin\theta_\mathrm{cam}C_\mathrm{TM}^-J_1(\gamma)\cos\psi\sqrt{\dfrac{\cos\theta_{\text{obj}}}{\cos\theta_\mathrm{cam}}}\sin2\theta_{\text{obj}}d\theta_{\text{obj}},
	\end{equation}
	\begin{equation}
	I_{zy}=\int_{0}^{\theta_{\text{obj}}^{\max}}i\sin\theta_\mathrm{cam}C_\mathrm{TM}^-J_1(\gamma)\sin\psi\sqrt{\dfrac{\cos\theta_{\text{obj}}}{\cos\theta_\mathrm{cam}}}\sin2\theta_{\text{obj}}d\theta_{\text{obj}},
	\end{equation}
	\begin{equation}
	I_{zz}=\int_{0}^{\theta_{\text{obj}}^{\max}}-2\sin\theta_\mathrm{cam}C_\mathrm{TM}^+J_0(\gamma)\sqrt{\dfrac{\cos\theta_{\text{obj}}}{\cos\theta_\mathrm{cam}}}\sin^2\theta_{\text{obj}}d\theta_{\text{obj}},
	\end{equation}
\end{subequations}
where $C_\mathrm{TE}=e^{iZ^+}+R^{\text{TE}}e^{iZ^-}$, $C_\mathrm{TM}^\pm =e^{iZ^+}\pm R^{\text{TM}}e^{iZ^-}$, $\gamma=\sqrt{\gamma_x^2+\gamma_y^2}$, $\psi=\tan^{-1}\dfrac{\gamma_y}{\gamma_x}$, $\psi\in[-\pi, \pi]$, and 
\begin{subequations}
	\begin{equation}
	\gamma_x = k_\mathrm{cam}\sin\theta_\mathrm{cam}x_\mathrm{cam}^o - k_{\text{obj}}\sin\theta_{\text{obj}} x_{\text{obj}}^\prime,
	\label{eq:expression_gammax}
	\end{equation}
	\begin{equation}
	\gamma_y = k_\mathrm{cam}\sin\theta_\mathrm{cam}y_\mathrm{cam}^o - k_{\text{obj}}\sin\theta_{\text{obj}} y_{\text{obj}}^\prime,
	\end{equation}
	\begin{equation}
	Z^\pm = k_\mathrm{cam}\cos\theta_\mathrm{cam}z_\mathrm{cam}^o \mp k_{\text{obj}}\cos\theta_{\text{obj}} z_{\text{obj}}^\prime.
	\end{equation}
	\label{eq:G_r_rp}
\end{subequations}
{\color{black}The identity $f_\mathrm{obj}\sin\theta_\mathrm{obj}=f_\mathrm{cam}\sin\theta_\mathrm{cam}$ has been used for the derivation.} The above integrals can be simplified by using the assumption $f_\mathrm{cam}\gg f_{\text{obj}}$ which yields the approximations $\sin\theta_\mathrm{cam}\approx 0$ and $\cos\theta_\mathrm{cam}\approx 1$ \cite{hu2011dyadic}. {\color{black} The lateral magnification and longitudinal magnification (assuming paraxial approximation and neglected reflections by the substrate interface) are derived in Supplement 1.}

\section{Investigations on resolution} 
Effects of the half space are studied by setting the refractive index of the lower half space as $n_\mathrm{sub}= 4.3$ (silicon) or $1.52$ (glass), while the objective region is with $n_{\text{obj}}=1.33$ (water) and the camera sensors are in the air, i.e., $n_\mathrm{cam}=1$. The wavelength in vacuum is chosen as $\lambda=$ \SI{500}{\nano\metre} and focal length $f_{\text{obj}}=$ \SI{1}{\centi\metre}, $f_\mathrm{cam}=$ \SI{10}{\centi\metre}. The lateral magnification is $M^\mathrm{lat}=13.3$. The substrate interface is positioned at the plane of $z_\mathrm{obj} = z^{b}_{\text{obj}}$ which is in the range of [\SI{-10}{\micro\metre},0]. Dipoles are assumed up to \SI{10}{\micro\metre} above the planar plane, i.e., $z_{\text{obj}}^\prime-z_{\text{obj}}^b\in$ [0,\SI{10}{\micro\metre}]. Where not stated explicitly, the NA of the system by default is assumed to be 1.

\subsection{Quantification of resolution}
\label{sec:quantification_resolution}
\begin{figure}[!ht]
	\centering
	\includegraphics[width=.46\linewidth]{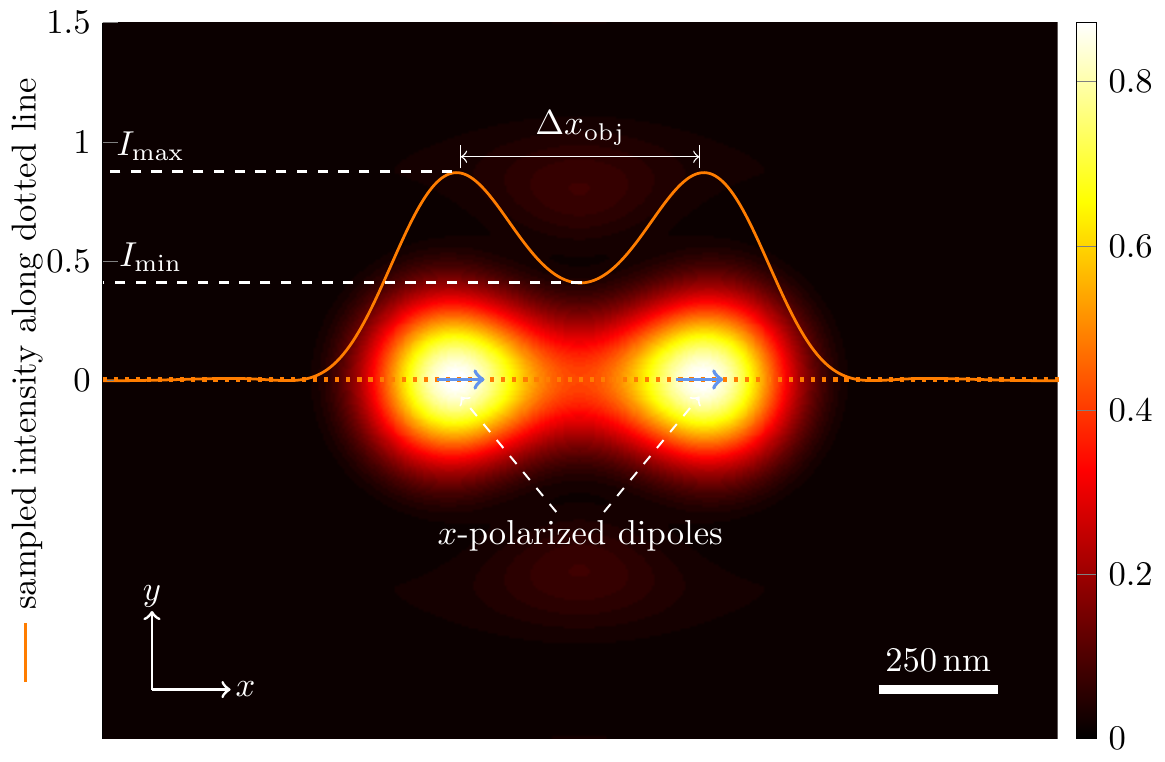}
	\caption{Image of two resolvable $x$-polarized dipoles when the silicon substrate is positioned  \SI{500}{\nano\metre} below the focal plane and the dipoles on the focal plane, resolution defined as the minimum of $\Delta x_\mathrm{obj}$ when $I_\mathrm{min}/I_\mathrm{max}\le 0.735$.}
	\label{fig:ExampleMap_xpol}	
\end{figure}

\begin{figure}[!ht]
	\centering
	\includegraphics[width=0.9\linewidth]{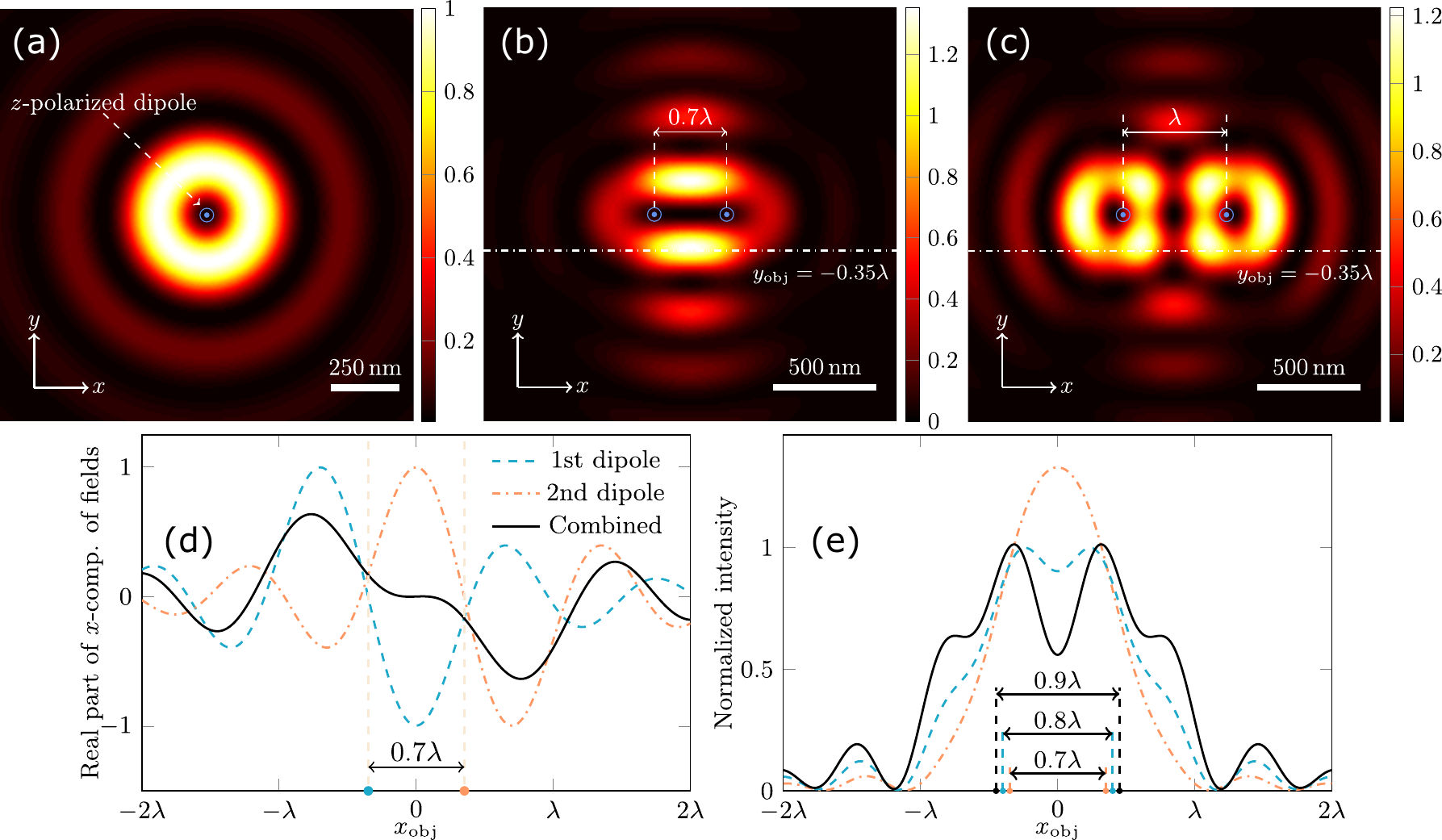}
	\caption{Observations of two $z$ polarized dipoles when the silicon substrate is \SI{500}{\nano\metre} below the focal plane and dipoles positioned at $[\pm\Delta x_{\mathrm{obj}}/2,0,0]$. {\color{black}(a) Image of a $z$-polarized dipole is an annulus. (b) Image of two unresolvable dipoles, small intensities between the dipoles due to destructive interference in (d). (c) Image of resolvable $z$-polarized dipoles. (e) Sampled intensities along the dashed lines in (b), (c) have decreased saddle-to-peak ratio with increasing distance between dipoles.} {\color{black} For the reference of the reader, the conventional rule of thumb value of resolution in the incoherennt case, given by Abbe is $\lambda/(2\mathrm{NA}) = $ \SI{250}{\nano\meter}.}}
	\label{fig:defRes_zpol_zpol}
\end{figure}

\begin{figure}[!ht]
	\centering
	\includegraphics[width=0.85\linewidth]{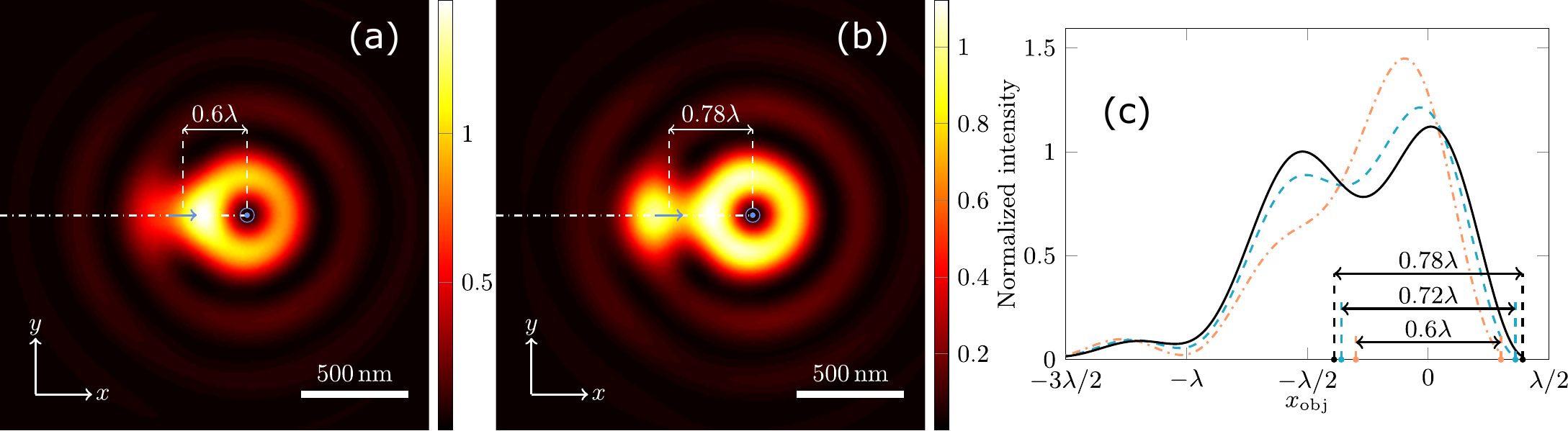}
	\caption{(a,b) Observations of the situation when there is a $x$ and a $z$-polarized dipole, the silicon substrate is \SI{500}{\nano\metre} below the focal plane and the dipoles positioned at $[\pm\Delta x_{\mathrm{obj}}/2,0,0]$. {\color{black}(c) Sampling intensities along the dash-dotted lines in (a), (b), the decreased saddle-to-peak ratio with increasing distance between two dipoles indicates the applicability of modified Rayleigh limit for quantifying resolution.}}
	\label{fig:defRes_xpol_zpol}
\end{figure}

The (induced) charge of dipoles can have a high variance in the analysis of imaging performance \cite{chen2012complete}. To get rid of effects from values of charge, normalization is performed. Considering electric fields at points $\vec{r}_{\mathrm{cam}}^{(i)},i=1,\ldots,N$, the normalized electric fields equal $E(\vec{r}_{\mathrm{cam}}^{(i)})/E^{\mathrm{max}}$, $E^{\mathrm{max}}=\max\{|E(\vec{r}_{\mathrm{cam}}^{(i)})|:i=1,\ldots,N\}$ and the related quantities are denoted by putting a bar above, i.e., $\bar{E}=E/E^{\mathrm{max}}$. The normalized field due to two dipoles is computed as $\bar{E}(\vec{r}_\mathrm{cam}^{(i)},\vec{r}_\mathrm{obj}^{1,\prime})+\bar{E}(\vec{r}_\mathrm{cam}^{(i)},\vec{r}_\mathrm{obj}^{2,\prime})$, where $\bar{E}(\vec{r}_\mathrm{cam}^{(i)},\vec{r}_\mathrm{obj}^{s,\prime})$ denotes the normalized field due to the $s$-th dipole. Thus, the intensity of combined fields may have maximum not equal to $1$. The silicon substrate is used in the example settings in Figs. \ref{fig:ExampleMap_xpol}-\ref{fig:defRes_xpol_zpol}.

\textbf{\color{black}Lateral resolution for x-polarized or y-polarized dipoles: }Imaging resolution is investigated from the intensity values in the image region when two dipoles are located near the focal point $O_{\text{obj}}$. Based on the image of two adjacent dipoles, as sketched by Fig. \ref{fig:ExampleMap_xpol}, the saddle-to-peak ratio $I_\mathrm{min}/I_\mathrm{max}$ is computed, where $I_\mathrm{max}$ is the maximum intensity of the combined field and $I_\mathrm{min}$ the minimum between two peaks. According to the modified Rayleigh limit, the resolution is defined as the minimum distance between two dipoles when the ratio is $\le0.735$. 

\textbf{\color{black}Lateral resolution for z-polarized dipoles: }The aforementioned criterion is based on the assumption that the image of a dipole is spot-like. This is true for $x$-polarized and $y$-polarized dipoles. However, as shown by Fig.~\ref{fig:defRes_zpol_zpol}(a), the image of a $z$-polarized dipole is an annulus. Then, observations along the $x$ or $y$ axis have two peaks, which make the above method for quantifying resolution inapplicable. We consider potential solution for defining resolution in the situation where at least one of the dipoles is z-polarized. Two cases are considered, two dipoles directed towards $\hat{z}$ and one of them directed to $\hat{x}$. We note that $y$-polarized dipoles are not considered due to the similar behaviors with $x$-polarized dipoles and straightforward generalizability..

Two dipoles are placed at $(-\Delta x_\mathrm{obj}/2,0,0)$ and $(\Delta x_\mathrm{obj}/2,0,0)$, respectively. In Fig.~\ref{fig:defRes_zpol_zpol}(b), we consider the situation when two $z$-polarized dipoles are separated by distance $\Delta x_{\mathrm{obj}}=0.7\lambda$ (i.e., \SI{350}{\nano\meter}). {\color{black}Even though the image of each dipole independently looks like annulus (see Fig.~\ref{fig:defRes_zpol_zpol}(a)), the image of two dipoles in vicinity does not look like a superimposition of two annulus (see Fig.~\ref{fig:defRes_zpol_zpol}(b).} Small intensities in the region between the dipoles are due to destructive interferences. Along the $x$ axis, using the DGF we determined that the intensity value is mainly contributed from the real part of the $x$-component of electric fields, denoted by $E_x$, which is plotted in Fig.~\ref{fig:defRes_zpol_zpol}(d). One sees that destructive interference occurs due to the opposite signs of $\Re({E}_x)$ between the dipoles. 

As the distance $\Delta x_\mathrm{obj}$ is increased, the signature of two dipoles becomes quite evident even though the image still does not look like a simple super-imposition of two annuli. Fig.~\ref{fig:defRes_zpol_zpol}(c) shows the image when $\Delta x_\mathrm{obj} = \lambda$. Through observations in Fig.~\ref{fig:defRes_zpol_zpol}, we propose that the resolvability of two $z$-polarized dipoles separated along x-axis can be inferred using the intensity profile along the line defined by $z_\mathrm{cam}=0$ and $y_{\mathrm{cam}}=\pm y_{\mathrm{cam}}^\mathrm{peak}$, as shown by the dash-dotted lines in Fig.~\ref{fig:defRes_zpol_zpol}(b,c). $y_{\mathrm{cam}}^\mathrm{peak}$ is the distance of peak value of intensities along $y$ axis to the center of the annulus and can be numerically determined. Fig.~\ref{fig:defRes_zpol_zpol}(e) shows the variation of the sampled intensities when the two dipoles are moved further from each other along this line for the example configuration used in this paper. The decreased saddle-to-peak ratio with the increasing distance between the emitters reveals that the modified Rayleigh limit computed over the line defined by $z_\mathrm{cam}=0$ and $y_{\mathrm{cam}}=\pm y_{\mathrm{cam}}^\mathrm{peak}$ can now be used for quantifying the resolution for the case of two $z$-polarized dipoles. 

\textbf{\color{black}Lateral resolution for one x- and one z-polarized dipole: }Images of the situation when a $x$-polarized dipole and a $z$-polarized are presented in the sample region are shown in Fig.~\ref{fig:defRes_xpol_zpol}(a) and (b). As seen, the only indicator of the presence of two dipoles is the asymmetry in the intensity profile in Fig.~\ref{fig:defRes_xpol_zpol}(a) whereas the presence of two dipoles is evident from the intensity profile in Fig.~\ref{fig:defRes_xpol_zpol}(b). Considering that the image of a z-polarized dipole is an annulus and not a spot, the
observations for quantifying the resolution are taken along a line segment that starts at the center of the annulus and extends beyond the peak corresponding to the other dipole, such as shown using the dash-dotted lines in Fig.~\ref{fig:defRes_xpol_zpol}(a,b). This is contrary to considering a line segment that passes through both of the dipoles in the case of only x-polarized or y-polarized dipoles as well as considering a line that passes through the peak of annuli parallel to the locations of dipole in the case of two z-polarized dipoles. As shown in Fig.~\ref{fig:defRes_xpol_zpol}(c), as the distance $\Delta x_\mathrm{obj}$ increases, the initially merged peak splits into two peaks and the saddle-to-peak ratio decreases. The modified Rayleigh limit can therefore be applied. However, it is important to remark that the location of peaks of sampled intensities does not correspond to the accurate position of dipoles.

\subsection{Depth of field (DOF)}

\begin{figure}[!ht]
	\centering 
	\includegraphics[width=0.9\linewidth]{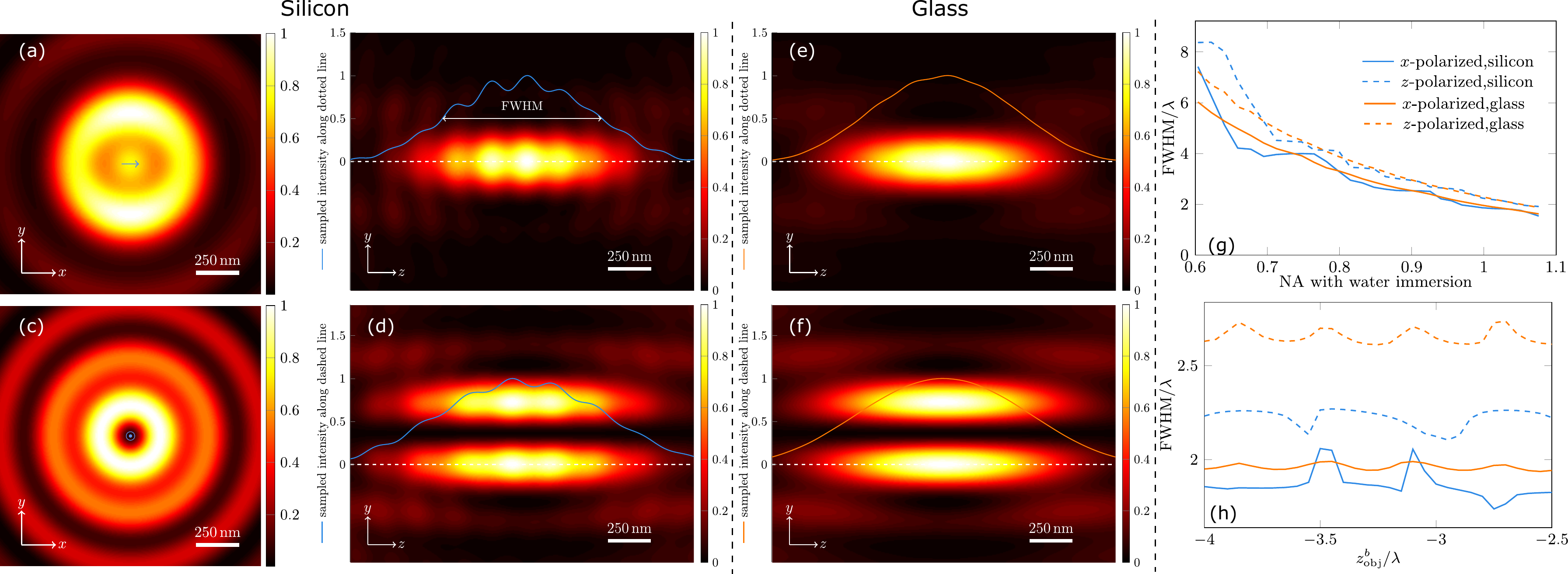}
	\caption{\color{black}With the silicon substrate \SI{2}{\micro\metre} below the focal plane, (a) gives an example of defocused image of the $x$-polarized dipole which is with coordinates $[0,0,1.5\lambda]$. (b) observations along $y$ axis with varied vertical positions of the dipole, the peak intensity tends to decreases when the dipole moves further from focal plane. (c,d) present results for the $z$-polarized dipole and (e,f) shows when the medium of substrate is glass. (g,h) show the variation of FWHM of the lines of peak values with increasing numerical aperture (NA) and varied position of substrate.}
	\label{fig:DefocusedMap}
\end{figure}

Depth of field is investigated by moving the position of a dipole along the optical axis. The substrate interface is assumed at the plane of $z_\mathrm{obj}^b=$\SI{-2}{\micro\metre}. As shown in Fig.~\ref{fig:ExampleMap_xpol} and \ref{fig:ExampleMap_zpol}, the image of a dipole at the focal point is a focused spot or an annulus with small side lobes. When the dipoles has the distance of $1.5\lambda$ to the focal plane, as presented in Fig.~\ref{fig:DefocusedMap}(a,c), the images become out of focus with high side lobes. We note also that the peak intensity is decreased as light spreads wider. Sampling intensities along the $y$ axis (since higher slide lobes observed than along the $x$ axis for the $x$-polarized dipole ), Fig.\ref{fig:DefocusedMap}(b,d) are obtained by varying the position of dipole {\color{black}with the silicon substrate, while (e,f) present the results with the glass substrate}. Despite the polarization and the substrate material, {\color{black}the peak values, which are the intensities along the dashed lines, tend to decrease when the dipole moves further from the focal plane}. Since small peak values indicate defocused images, FWHM (full width at half maximum) of the line graphs is used to qualitatively measure the depth of field. The above results are with NA$=1$. Keeping the immersion medium as water, the effects of numerical aperture is studied by varying the angular semiaperture. As seen from Fig.~\ref{fig:DefocusedMap}(g), the DOF is smaller with the increasing NA. We also note that even though the DOF is usually reported as a single number, the sensitivity to the polarization of the dipole is not often reported such as done here. The variation of DOF when using glass or silicon substrate is not significant in Fig.~\ref{fig:DefocusedMap}(g). {\color{black}The $x$-polarized dipole generally have smaller focused sizes than the $z$-polarized dipole. Fig. 6(h) shows the sensitivity of the DOF to the location of the distances of the interface from the focus. We note a certain oscillatory behaviour of the value of DOF, which is more prominent in the case of silicon as compared to the glass substrate.}  

\subsection{Comparison of lateral resolution for silicon and glass substrate.}
Since $x$-polarized and $y$-polarized dipoles behave similarly (see Supplement 1), we focus on the 
$4$ representative cases  for studying lateral resolution. These are with polarizations
$\{\vec{p}_1=\hat{x}, \vec{p}_2=\hat{x}\}$, $\{\vec{p}_1=\hat{x}, \vec{p}_2=\hat{y}\}$, $\{\vec{p}_1=\hat{z}, \vec{p}_2=\hat{z}\}$, and
$\{\vec{p}_1=\hat{x}, \vec{p}_2=\hat{z}\}$, where $\vec{p}_i$ is the direction of the $i$-th dipole, $i=1,2$. The second case is due to the fact that it may have significantly higher resolution than the first case (see Supplement 1). 

Setting the positions of dipoles are $(-\Delta x_\mathrm{obj}/2,0,z_\mathrm{obj}^\prime)$ and $(\Delta x_\mathrm{obj}/2,0,z_\mathrm{obj}^\prime)$, the modified Rayleigh limit is obtained by increasing $\Delta x_\mathrm{obj}$ with the step $0.02\lambda$ and taking the minimal $\Delta x_\mathrm{obj}$ when the saddle-to-peak ratio is $\le0.735$. We assume that two dipoles have the distance to the focal plane less than or equal to $\lambda$ and the substrate interface is below the focal plane up to $2\lambda$, i.e. $z^b_\mathrm{obj} \in (0,-2\lambda]$. {\color{black}The results without the presence of half space are also included for comparisons}.

\begin{figure}[!ht]
	\centering 
	\includegraphics[width=0.95\linewidth]{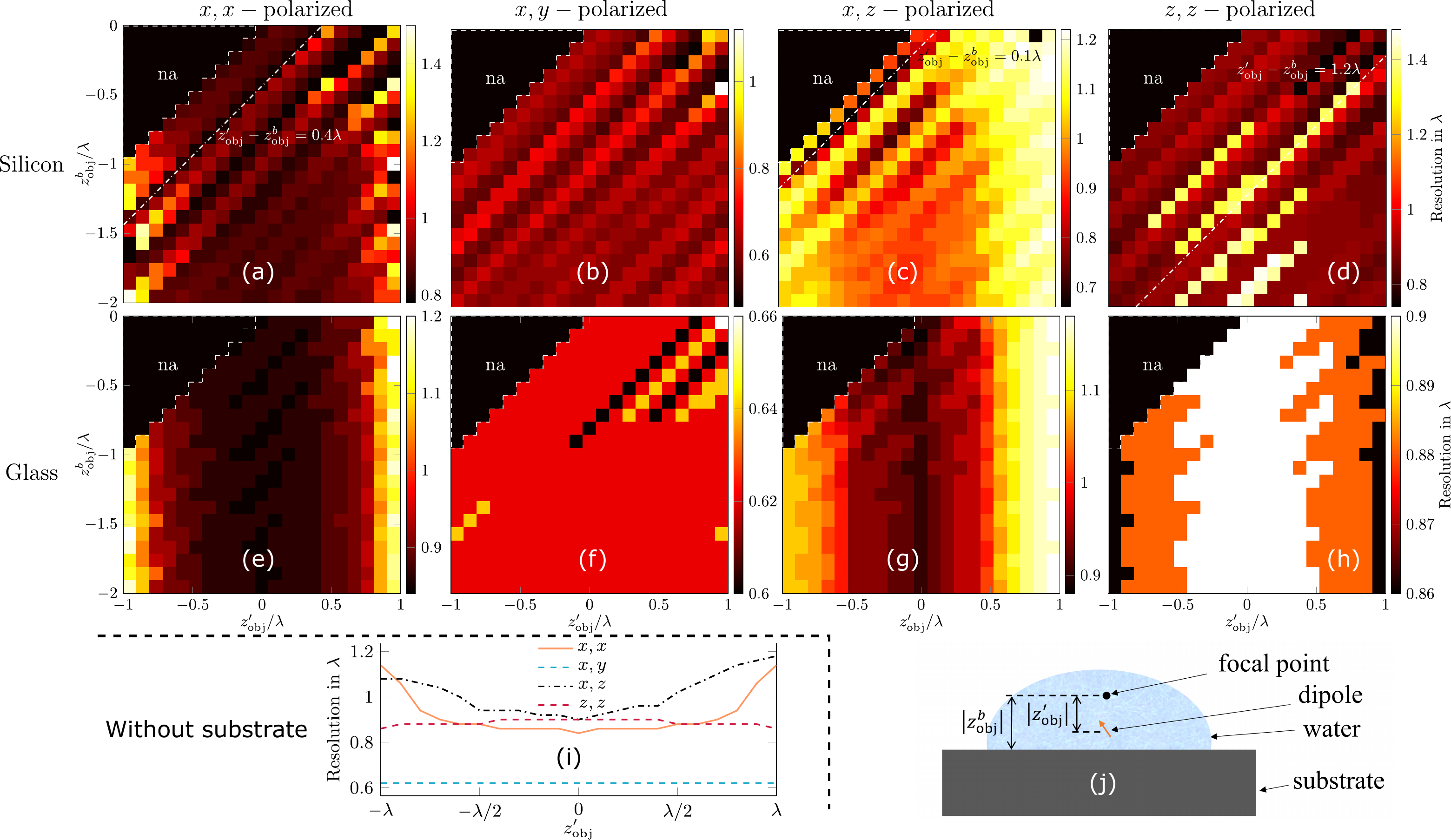}
	\caption{\color{black}Study of lateral resolution by varying the position of dipoles and the substrate, undefined cases (with ``na") when dipoles are below the substrate. Figures in the 1st row are obtained with the silicon graphite and those in the 2nd row are with the glass substrate. Four sets of dipole polarization are considered for each configuration. Cases on the dashed lines are examples of the observation that the lateral resolution is much impacted by the distance between the dipole and the substrate surface. (i) gives results without substrate. (j) explains the involved notations.}
	\label{fig:lateralResolution}
\end{figure}

\begin{figure}[!ht]
	\centering 
	\includegraphics[width=0.95\linewidth]{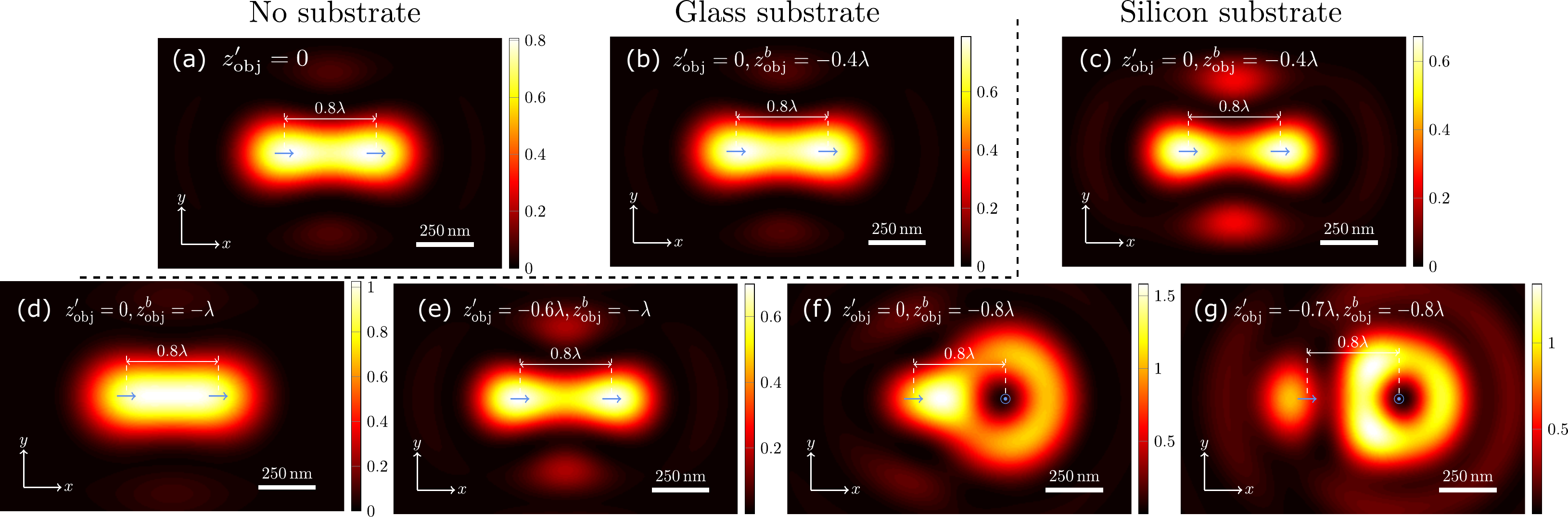}
	\caption{\color{black}(a-c) present the best case for the three situations when the $x$-polarized dipoles are separated by distance $0.8\lambda$. Higher resolution is reached with the presence of silicon substrate in the sample region. (c,d) give an example of effects from the position of the silicon substrate, while (d,e) and (f,g) are examples for showing the effects from the vertical position of the dipoles.}
	\label{fig:exampleLateralRes}
\end{figure}

The estimated resolution in Fig.~\ref{fig:lateralResolution} is quantified in $\lambda$. The region bounded by white dashed lines and filled with black represents the non-physical condition when the dipoles are below the substrate interface and therefore not applicable for study. {\color{black}The estimated resolutions with the glass substrate are presented in Fig.~\ref{fig:lateralResolution}(e-h). Due the small refractive index difference with water, the variation of resolution follows the similar phenomena with the situation when no presence of the substrate (the associated results in Fig.~\ref{fig:lateralResolution}(i)). The common observations consistent between the no-substrate and glass-substrate are presented here. In particular, we report the observations related to the resolution in the four cases when the $z$-location of the dipoles is changed. The resolution tends to be better when the two $x$-polarized dipoles and the $x,z$-polarized dipoles get closer to the focal plane. For the two $z$-polarized dipoles, the estimated resolutions range from $0.86\lambda$ to $0.9\lambda$. The resolution for the $x,y$-polarized dipoles is a constant, except the cases which provide strong evidences for the dependency on the distance between the dipoles and the substrate interface. Moreover, for all considered settings of $z_\mathrm{obj}^\prime$ and $z_\mathrm{obj}^b$, the $x,y$-polarized two dipoles are imaged with the best resolution.
	
	As seen from Fig.~\ref{fig:lateralResolution}(a-d), the above-mentioned traits also appear with the silicon substrate. Examples are indicated by the dashed lines in Fig.~\ref{fig:lateralResolution}(a,c,d). However, the high contrast of silicon (relative to water) leads to more complex trends. For instance, the resolution of the two $x$-polarized dipoles only has small variations when $\abs{z_\mathrm{obj}^\prime}\le0.4\lambda$, but is very sensitive to the position of substrate when the dipoles are moved further. Irrespective of the dipole polarization, the microscope with the silicon substrate shows a larger range of resolution than with the glass substrate. Interestingly, it can achieve higher resolutions for the best cases. Examples are given in Fig.~\ref{fig:exampleLateralRes}(a-c), where the two $x$-polarized dipoles are not well separated in the images without the substrate or with the glass substrate but are clearly distinguishable using the silicon substrate.}

The estimations in Fig.~\ref{fig:lateralResolution} also reveal that samples on the focal plane may have lower-resolution images than those off the plane depending upon the position of the interface. Fig.~\ref{fig:exampleLateralRes}(d,e) show that the non-resolvable $x$-polarized dipoles on the focal plane becomes resolvable when the dipoles are on the plane $0.6\lambda$ lower. Fig.~\ref{fig:exampleLateralRes}(f,g) present the similar phenomenon for the $x,z$-polarized dipoles. However, it is should be noted that the shift of dipoles off the focal plane may lead to higher side lobes, as seen in Fig.~\ref{fig:exampleLateralRes}(g). 

\subsection{\color{black}Longitudinal magnification}
\begin{figure}[!ht]
	\centering 
	\includegraphics[width=0.9\linewidth]{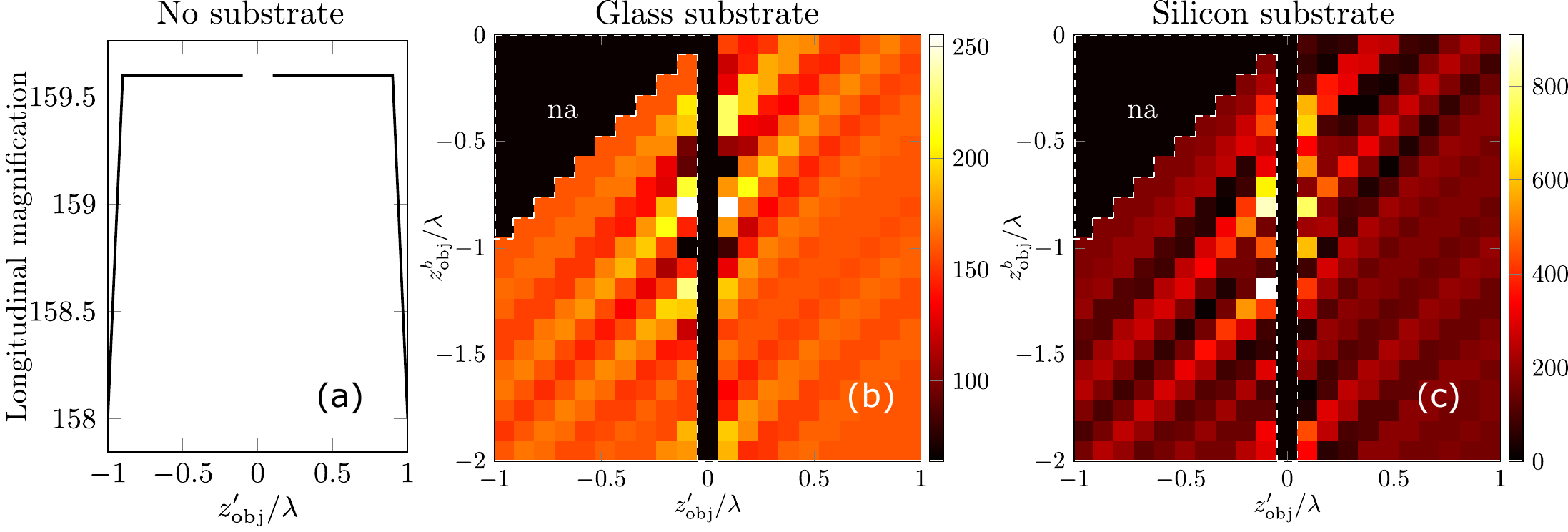}
	\caption{\color{black} (a)When NA$=1$ and $\abs{z_\mathrm{obj}^\prime}\le 0.9\lambda$, the longitudinal magnification is estimated as $159.6$. (b) Putting the glass substrate, the estimated magnification has strong fluctuations with the varied position of the substrate when $\abs{z_\mathrm{obj}^\prime}\le 0.2\lambda$. Undefined cases due to settings that the dipoles are below the substrate interface or $z_\mathrm{obj}^\prime=0$. (c) Estimations when the medium of substrate is silicon.}
	\label{fig:longitudinalMag}
\end{figure}
Supplement 1 provides the derivation of lateral and longitudinal magnification. The general derivation of longitudinal resolution employs paraxial approximation and neglects reflections from the substrate, which may be not valid. Based on the derived formula of Eq.~(S3), we have the longitudinal magnification $M^\mathrm{lon}=133$ for the concerned situation, i.e., water immersion and $f_\mathrm{cam}/f_{\text{obj}}=10$. However, this estimation is strongly deviant from the numerical estimations in Fig.~\ref{fig:longitudinalMag}, which are obtained by putting a $x$-polarized dipole at $[0,0,z_\mathrm{obj}^\prime]$ and identifying the location of maximum intensity. Since NA$=1$, the paraxial approximation is invalid. Without the substrate, the estimations show that $M^\mathrm{lon}$ is a constant and equals $159.6$ when $\abs{z_\mathrm{obj}^\prime}\le 0.9\lambda$. The presence of the substrate has great impact on the longitudinal magnification. With the glass substrate, strongly fluctuated estimations are obtained when the dipole is close to the focal plane and the magnification ranges from $63.84$ to $255.36$. With the silicon substrate, the variation extent is even larger.

\subsection{Comparison of longitudinal resolution for silicon and glass substrates.}

\begin{figure}[!ht]
	\centering 
	\includegraphics[width=\linewidth]{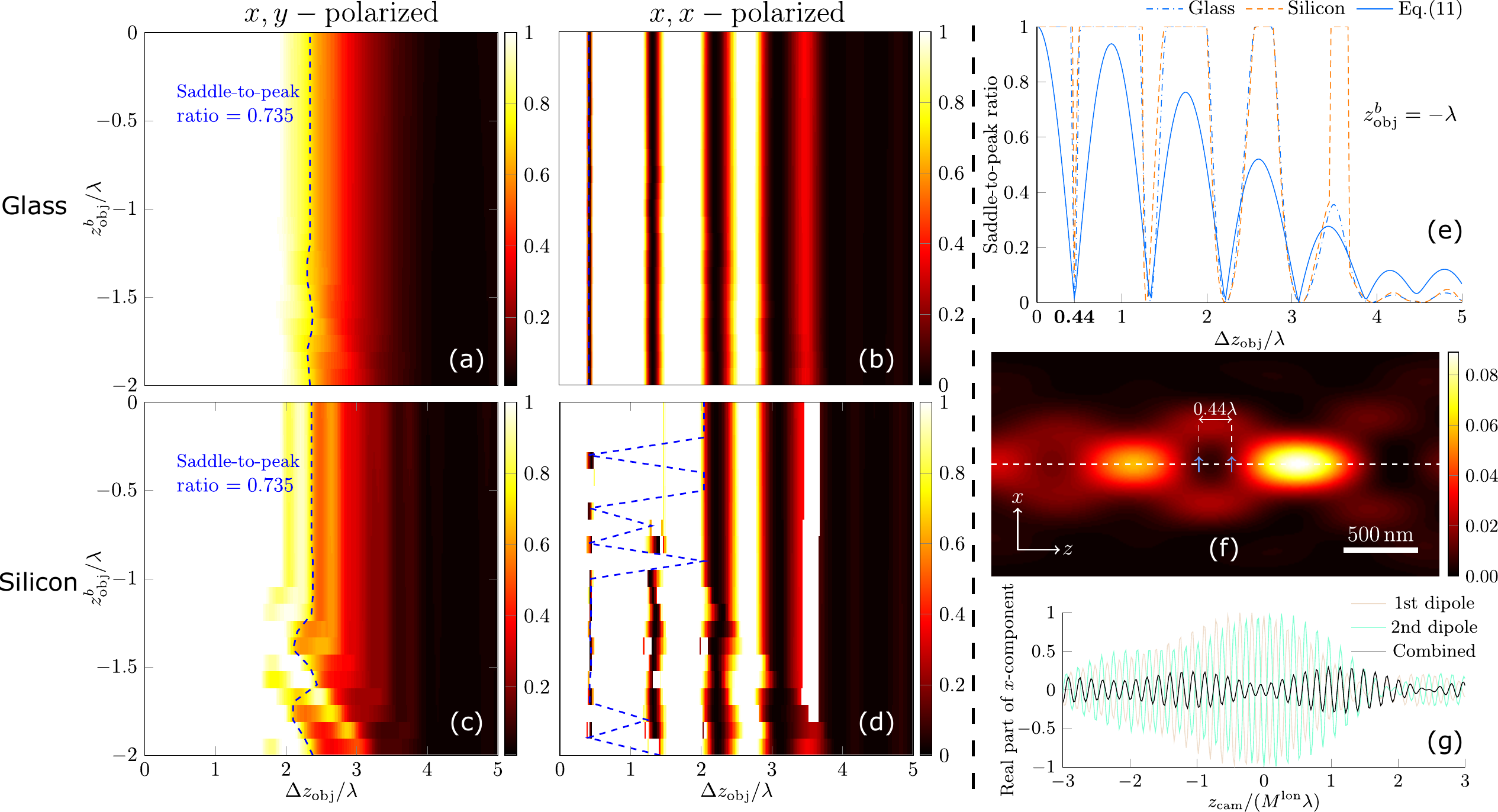}
	\caption{\color{black}(a-d) Variation of saddle-to-peak ratio with increasing distance between the two dipoles and different positions of the substrate interface. The dashed lines indicate the modified Rayleigh limit. (e) The oscillatory behaviors of saddle-to-peak ratio when the two dipoles are directed to $\hat{x}$ may be explained by the observation of intensity at the focal point in the image region. (f,g) reveal that the first minimum in (e) is due to destructive inference and the spot locations mismatch the positions of the dipoles. Longitudinal magnification $M^\mathrm{lon}=206$ is used for illustration.}
	\label{fig:longitudinalRes}
\end{figure}

The longitudinal resolution is studied by putting two dipoles at the optical axis but separated by distance $\Delta z_\mathrm{obj}$, which is increased from $0$ by the step $0.02\lambda$. When $z_\mathrm{obj}^b<-\Delta z_\mathrm{obj}$, the two dipoles are assumed with $z$ coordinates $-\Delta z_\mathrm{obj}/2$ and $\Delta z_\mathrm{obj}/2$, respectively. Otherwise, the $z$ coordinates equal $z_\mathrm{obj}^b$ and $z_\mathrm{obj}^b+\Delta z_\mathrm{obj}$ to ensure two dipoles are above the substrate. Fig.~\ref{fig:longitudinalRes}(a-d) show the variation of saddle-to-peak ratio with the increasing distance between dipoles and the varied position of the substrate. The ratio would be set as $1$ when the two dipoles cannot be clearly distinguished (e.g., images of a single spot or with high side lobes). As seen, for the $x,y$-polarized dipoles, the ratio monotonically decreases as the distance between dipoles is larger with the glass substrate. However, for the $x,x$-polarized dipoles, the saddle-to-peak ratio shows oscillatory behaviors which can make the two dipole unresolvable even when the distance is larger than the modified Rayleigh limit, which is determined by the threshold of $0.735$. 

The oscillatory behaviors have been reported before \cite{chen2012resolution,sheppard1994confocal} and explained by observing the intensity at the focal point of the image region $O_\mathrm{cam}$, which is a saddle point for situations when the dipoles locate symmetrically to the focal plane in the sample region. Therefore, the intensity at $O_\mathrm{cam}$ is proportional to the value of saddle-to-peak ratio. Since $O_\mathrm{cam}$ and the dipoles are on the optical axis, we have $\gamma=0$ and $z_\mathrm{cam}=0$ in Eq.~\eqref{eq:solG}. Since $J_1(0)=J_2(0)=0$, only $I_{xx}^{(1)} \ne 0$ for the field solution to the $x$-polarized dipoles. Letting $z_\mathrm{obj}^\prime=\pm \Delta z_\mathrm{obj}/2$, the expression of $I_{xx}^{(1)}$ is rewritten as
\begin{equation}
\label{eq:combineIxx1}
I_{xx}^{(1)} =\int_{0}^{\theta_{\text{obj}}^{\max}}A(\theta_{\text{obj}})e^{ik_\mathrm{cam}\cos\theta_\mathrm{cam}z_\mathrm{cam}}\cos\bracket{k_\mathrm{obj}\cos\theta_{\text{obj}}\Delta z_\mathrm{obj}/2}d\theta_{\text{obj}},
\end{equation}
where
\begin{equation}
A(\theta_{\text{obj}}) = 2\sbracket{1+R^{\text{TE}}+\cos\theta_\mathrm{cam}\cos\theta_{\text{obj}}(1-R^{\text{TM}})}\sqrt{\dfrac{\cos\theta_{\text{obj}}}{\cos\theta_\mathrm{cam}}}\sin\theta_{\text{obj}}.
\end{equation}
Setting $z_\mathrm{cam}=0$, $I_{xx}^{(1)}$ is a function of 
$\Delta z_\mathrm{obj}$ and the normalized $I_{xx}^{(1)}$ is plotted in Fig.~\ref{fig:longitudinalRes}(e). The local minima coincide with the minima of the line graphs of saddle-to-peak ratio for the three situations, i.e., without substrate and the presence of glass substrate and silicon substrate \SI{500}{\nano\meter} below the focal plane.   

{\color{black}Fig.~\ref{fig:longitudinalRes}(f) shows the image of the two $x$-polarized dipoles, the locations of which correspond with the first minimum in Fig.~\ref{fig:longitudinalRes}(e). While two spots are exhibited, the peak value is very small and the positions of spots mismatch the positions of dipoles. This phenomenon is explained by observations of the sampled electric fields along the dashed line. Fig.~\ref{fig:longitudinalRes}(g) describes the real part of the $x$ component of electric fields. As seen, responses due to the two dipoles interfere destructively. In consequence, a minimum amplitude is observed at $O_\mathrm{cam}$, which leads to the resolvability of the two dipoles, but the positions of the peak values are shifted.} 

\section{Conclusion}
The dyadic Green's function (DGF) is solved as multiple Sommerfeld integrals. Our formulation includes the effect of reflections from the substrate over which the sample rests as well as realistically high numerical aperture for high NA water immersion systems. This allows us to emulate the 3D full-field effects and thereby more accurately study the resolution and the other effects, especially as a consequence of using high refractive index substrate such as silicon.

The lateral resolution is found dependent on the polarization of dipoles and also the position of the substrate interface in the case of silicon substrate. It is also demonstrated that such variation can be neglected when using glass substarte because of small refractive index contrast with water medium. It is also noted that simply using a high refractive index substrate can alter the range of resolution significantly. However, the effect of high refractive index substrate on DOF is not quite significant. 

The longitudinal resolution is studied with two cases, $\{\vec{p}_1, \vec{p}_2=\hat{x}\}$ and $\{\vec{p}_1=\hat{x}, \vec{p}_2=\hat{y}\}$. In the former case, as the distance between dipoles increases, the saddle-to-peak ratio has oscillatory behaviors which can make the two dipole unresolvable even when the distance is larger than the conventional definition of resolution. In the latter case, {\color{black}with the glass substrate, the ratio monotonically decreases as the distance between dipoles is larger. However, it is not always true with the silicon substrate.}

While this study indicate the possibility of achieving better resolution using silicon substrate, it also provides insight into large sensitivity of image behaviour on the relative position of interface and dipoles. We expect that this analysis and the full-field derivation of DGF will open new avenues for exploring high refractive index materials as substrates for use in coherent microscopy. 

\bibliographystyle{chicago}
\bibliography{OE-Liu-etal}

\begin{thebibliography}{}

\bibitem[\protect\citeauthoryear{Abubakar, Van~den Berg, and
  Mallorqui}{Abubakar et~al.}{2002}]{abubakar2002imaging}
Abubakar, A., P.~M. Van~den Berg, and J.~J. Mallorqui (2002).
\newblock Imaging of biomedical data using a multiplicative regularized
  contrast source inversion method.
\newblock {\em IEEE Trans. Microw. Theory Techn.\/}~{\em 50\/}(7), 1761--1771.

\bibitem[\protect\citeauthoryear{Agarwal and Mach{\'a}{\v{n}}}{Agarwal and
  Mach{\'a}{\v{n}}}{2016}]{agarwal2016multiple}
Agarwal, K. and R.~Mach{\'a}{\v{n}} (2016).
\newblock Multiple signal classification algorithm for super-resolution
  fluorescence microscopy.
\newblock {\em Nat. Commun.\/}~{\em 7\/}(1), 1--9.

\bibitem[\protect\citeauthoryear{Ahmad, Dubey, Butola, Tinguely, Ahluwalia, and
  Mehta}{Ahmad et~al.}{2020}]{ahmad2020sub}
Ahmad, A., V.~Dubey, A.~Butola, J.-C. Tinguely, B.~S. Ahluwalia, and D.~S.
  Mehta (2020).
\newblock Sub-nanometer height sensitivity by phase shifting interference
  microscopy under environmental fluctuations.
\newblock {\em Opt. Express\/}~{\em 28\/}(7), 9340--9358.

\bibitem[\protect\citeauthoryear{Archetti, Glushkov, Sieben, Stroganov,
  Radenovic, and Manley}{Archetti et~al.}{2019}]{archetti2019waveguide}
Archetti, A., E.~Glushkov, C.~Sieben, A.~Stroganov, A.~Radenovic, and S.~Manley
  (2019).
\newblock Waveguide-paint offers an open platform for large field-of-view
  super-resolution imaging.
\newblock {\em Nat. Commun.\/}~{\em 10\/}(1), 1--9.

\bibitem[\protect\citeauthoryear{Chen, Agarwal, Sheppard, Phang, and Chen}{Chen
  et~al.}{2012}]{chen2012resolution}
Chen, R., K.~Agarwal, C.~J. Sheppard, J.~C. Phang, and X.~Chen (2012).
\newblock Resolution of aplanatic solid immersion lens based microscopy.
\newblock {\em J. Opt. Soc. Am. A\/}~{\em 29\/}(6), 1059--1070.

\bibitem[\protect\citeauthoryear{Chen, Agarwal, Sheppard, Phang, and Chen}{Chen
  et~al.}{2013}]{chen2013complete}
Chen, R., K.~Agarwal, C.~J. Sheppard, J.~C. Phang, and X.~Chen (2013).
\newblock A complete and computationally efficient numerical model of aplanatic
  solid immersion lens scanning microscope.
\newblock {\em Opt. Express\/}~{\em 21\/}(12), 14316--14330.

\bibitem[\protect\citeauthoryear{Chen, Agarwal, Zhong, Sheppard, Phang, and
  Chen}{Chen et~al.}{2012}]{chen2012complete}
Chen, R., K.~Agarwal, Y.~Zhong, C.~J. Sheppard, J.~C. Phang, and X.~Chen
  (2012).
\newblock Complete modeling of subsurface microscopy system based on aplanatic
  solid immersion lens.
\newblock {\em J. Opt. Soc. Am. A\/}~{\em 29\/}(11), 2350--2359.

\bibitem[\protect\citeauthoryear{Crocco, Catapano, Di~Donato, and
  Isernia}{Crocco et~al.}{2012}]{crocco2012linear}
Crocco, L., I.~Catapano, L.~Di~Donato, and T.~Isernia (2012).
\newblock The linear sampling method as a way to quantitative inverse
  scattering.
\newblock {\em IEEE Trans. Antennas Propag.\/}~{\em 60\/}(4), 1844--1853.

\bibitem[\protect\citeauthoryear{Gibson and Lanni}{Gibson and
  Lanni}{1991}]{gibson1991experimental}
Gibson, S.~F. and F.~Lanni (1991).
\newblock Experimental test of an analytical model of aberration in an
  oil-immersion objective lens used in three-dimensional light microscopy.
\newblock {\em J. Opt. Soc. Am. A\/}~{\em 8\/}(10), 1601--1613.

\bibitem[\protect\citeauthoryear{Guo, Zhuang, Chen, and Liang}{Guo
  et~al.}{2007}]{guo2007multilayered}
Guo, H., S.~Zhuang, J.~Chen, and Z.~Liang (2007).
\newblock Multilayered optical memory with bits stored as refractive index
  change. {I. Electromagnetic theory}.
\newblock {\em J. Opt. Soc. Am. A\/}~{\em 24\/}(6), 1776--1785.

\bibitem[\protect\citeauthoryear{Hu, Chen, Agarwal, Sheppard, Phang, and
  Chen}{Hu et~al.}{2011}]{hu2011dyadic}
Hu, L., R.~Chen, K.~Agarwal, C.~J. Sheppard, J.~C. Phang, and X.~Chen (2011).
\newblock Dyadic green’s function for aplanatic solid immersion lens based
  sub-surface microscopy.
\newblock {\em Opt. Express\/}~{\em 19\/}(20), 19280--19295.

\bibitem[\protect\citeauthoryear{Kong}{Kong}{2003}]{GreenKong}
Kong, J.~A. (2003).
\newblock Green’s functions for planarly layered media.
\newblock Technical report, Massachusetts Institute of Technology.

\bibitem[\protect\citeauthoryear{Marx}{Marx}{2019}]{Marx2019It}
Marx, V. (2019).
\newblock It’s free imaging — label-free, that is.
\newblock {\em Nat. Methods\/}~{\em 16}, 1209–1212.

\bibitem[\protect\citeauthoryear{Novotny and Hecht}{Novotny and
  Hecht}{2012}]{novotny2012principles}
Novotny, L. and B.~Hecht (2012).
\newblock {\em Principles of nano-optics}.
\newblock Cambridge university press.

\bibitem[\protect\citeauthoryear{Sheppard, Connolly, Lee, and
  Cogswell}{Sheppard et~al.}{1994}]{sheppard1994confocal}
Sheppard, C.~J., T.~J. Connolly, J.~Lee, and C.~J. Cogswell (1994).
\newblock Confocal imaging of a stratified medium.
\newblock {\em Appl. Opt.\/}~{\em 33\/}(4), 631--640.

\bibitem[\protect\citeauthoryear{Stockert and Bl{\'a}zquez-Castro}{Stockert and
  Bl{\'a}zquez-Castro}{2017}]{stockert2017fluorescence}
Stockert, J.~C. and A.~Bl{\'a}zquez-Castro (2017).
\newblock {\em Fluorescence Microscopy in Life Sciences}.
\newblock Bentham Science Publishers.

\end{thebibliography}

\section*{Supplement 1}
\renewcommand{\theequation}{S.\arabic{equation}}
\setcounter{equation}{0}
\beginsupplement

{\bf Magnification:} The dependency of $\tensor{G}$ on the observation point $\vec{r}_\mathrm{ccd}^o$ and the dipole position $\vec{r}_\mathrm{ccd}^\prime$ is expressed in Eqs.~(10), from which the magnification can be derived. Assume the $x$ coordinate of the dipole has a shift $\Delta x_{\text{obj}}^\prime$. Since only $\gamma_x$ is related with $x_{\text{obj}}^\prime$ and
\begin{equation}
\gamma_x=k_\mathrm{ccd}\sin\theta_\mathrm{ccd}\bracket{x_\mathrm{ccd}^o -\dfrac{n_{\text{obj}}}{n_\mathrm{ccd}}\dfrac{f_\mathrm{ccd}}{f_{\text{obj}}} x_{\text{obj}}^\prime},
\label{eq:rewritten_expression_gammax}
\end{equation} 
$x^o_\mathrm{ccd}$ should have the shift $M^\mathrm{lat}\Delta x_{\text{obj}}^\prime$, $M^\mathrm{lat}=\dfrac{n_{\text{obj}}}{n_\mathrm{ccd}}\dfrac{f_\mathrm{ccd}}{f_{\text{obj}}}$, in order to keep the value of $\gamma_x$ unchanged. The magnification along the $y$ axis is similarly analyzed. $M^\mathrm{lat}$ is the so-called lateral magnification.

For the longitudinal magnification, when the scatterings from the planar interface can be neglected, i.e., $R^{\text{TM}}, R^{\text{TE}}\approx 0$, the terms related with $Z^-$ are treated as zero. With the paraxial approximation, i.e., $\theta_{\text{obj}}^\mathrm{max}\rightarrow 0$, we have approximations $\cos\theta_\mathrm{ccd}\approx1-0.5(f_{\text{obj}}/f_\mathrm{ccd})^2\sin^2\theta_{\text{obj}}$ and $\cos\theta_{\text{obj}}\approx1-0.5\sin^2\theta_{\text{obj}}$.
Thus,
\begin{equation}
\label{eq:approx_Zp}
Z^+=\bracket{k_\mathrm{ccd}z_\mathrm{ccd}^o-k_{\text{obj}}z^\prime_{\text{obj}}}-\dfrac{k_\mathrm{ccd}}{2}\bracket{\dfrac{f_{\text{obj}}}{f_\mathrm{ccd}}}^2\sin^2\theta_{\text{obj}}\bracket{z_\mathrm{ccd}^o-\dfrac{n_{\text{obj}}}{n_\mathrm{ccd}}\bracket{\dfrac{f_\mathrm{ccd}}{f_{\text{obj}}}}^2z_{\text{obj}}^\prime}.
\end{equation} 
From the integrals in Eq.~(9), we see that $Z^+$ impacts the solution through $e^{iZ^+}$. Since the first summation term of \eqref{eq:approx_Zp} is not related with $\theta_{\text{obj}}$, it can be moved outside from the integrals and has no impact on the amplitude of the integral. From the second summation term, the longitudinal magnification is derived as
\begin{equation}
\label{eq:longitudinalMagnification}
M^\mathrm{lon}=\dfrac{n_{\text{obj}}}{n_\mathrm{ccd}}\bracket{\dfrac{f_\mathrm{ccd}}{f_{\text{obj}}}}^2.
\end{equation}
Remark that the derivation of $M^\mathrm{lat}$ is with no assumptions or approximations, while the expression of $M^\mathrm{lon}$ is with the assumptions that $R^{\text{TM}}, R^{\text{TE}}\approx 0$ and $\theta_{\text{obj}}^\mathrm{max}\rightarrow 0$.

However, the assumption of paraxial approximation and neglected reflections by the interface may be invalid, especially when the substrate has a high contrast with the immersion medium.  

{\bf Properties of dyadic Green's function:} Only the resolution along the $x$ direction is analyzed based on the following study of ellipticity. Putting a $z$-polarized dipole at the optical axis, the observed intensity would be the same if the $x$ coordinate and the $y$ coordinate of the observation point are exchanged, i.e., $I(\vec{r}_\mathrm{ccd})=I(\vec{r}_\mathrm{ccd}^*)$, $\vec{r}_\mathrm{ccd}=[x_\mathrm{ccd},y_\mathrm{ccd},0]$ and $\vec{r}_\mathrm{ccd}^*=[y_\mathrm{ccd},x_\mathrm{ccd},0]$. This property can be derived from the solution of DGF in Eq.~(9). The observed intensity is the summation of intensity of three field components, i.e., $I = |I_{xz}|^2+|I_{yz}|^2+|I_{zz}|^2$. The observation position impacts the field solution through $\psi$ and $\gamma$. From the definitions in Eq.~(10), we see $\cos\psi(\vec{r}_\mathrm{ccd})=\sin\psi(\vec{r}_\mathrm{ccd}^*)$, $\sin\psi(\vec{r}_\mathrm{ccd})=\cos\psi(\vec{r}_\mathrm{ccd}^*)$ and $\gamma(\vec{r}_\mathrm{ccd})=\gamma(\vec{r}_\mathrm{ccd}^*)$. Consequently, $I_{xz}(\vec{r}_\mathrm{ccd})=I_{yz}(\vec{r}_\mathrm{ccd}^*)$, $I_{yz}(\vec{r}_\mathrm{ccd})=I_{xz}(\vec{r}_\mathrm{ccd}^*)$, and $I_{zz}(\vec{r}_\mathrm{ccd})=I_{zz}(\vec{r}_\mathrm{ccd}^*)$. Thus, $I(\vec{r}_\mathrm{ccd})=I(\vec{r}_\mathrm{ccd}^*)$. This identity reveals that no ellipticity for $z$-polarized dipoles. 

\begin{figure}[!ht]
	\captionsetup[subfigure]{justification=centering}
	\begin{subfigure}{.45\textwidth}
		\centering
		\includegraphics[width=\linewidth]{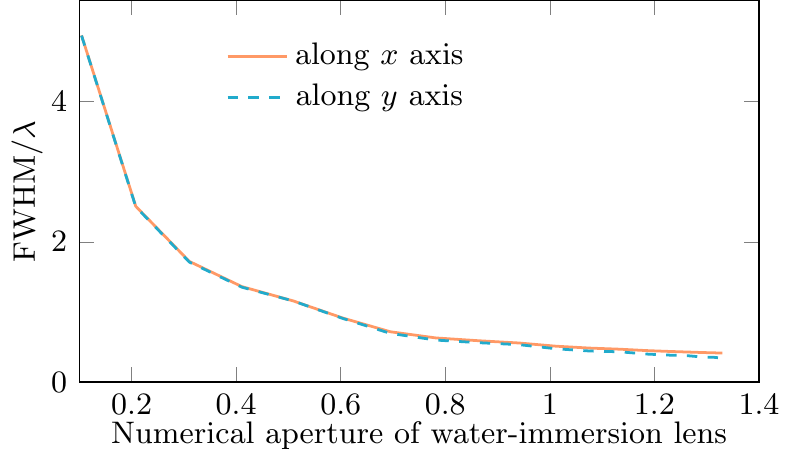}
		\caption{}
		\label{fig:ellipticity_vs_thetamax}
	\end{subfigure}
	\begin{subfigure}{.45\textwidth}
		\centering 
		\includegraphics[width=\linewidth]{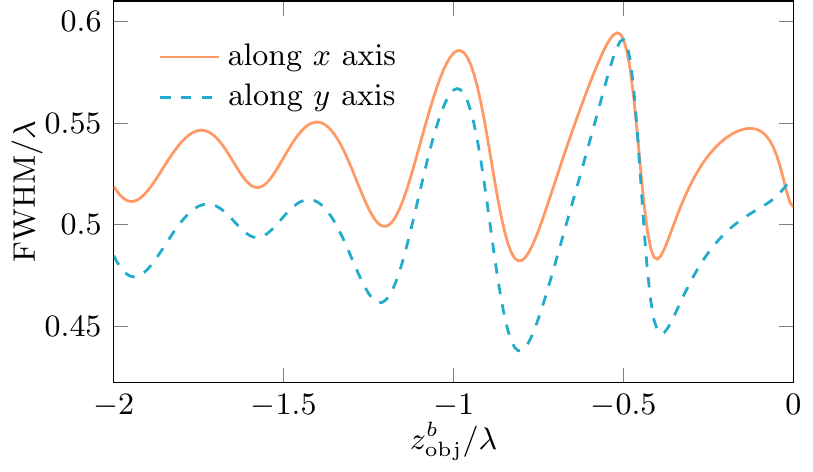}
		\caption{}
		\label{fig:ellipticity_vs_zobjb}
	\end{subfigure}
	\caption{Study of ellipticity with the $x$-polarized dipole by varying (a) the numerical aperture of the water-immersion objective lens when the silicon substrate is \SI{1}{\micro\metre} below the focal plane and (b) varying the position of the substrate when $\mathrm{NA}=1$.}
	\label{fig:Elliticity}
\end{figure}
To check the ellipticity for $x$-polarized dipoles, values of FWHM associated with observations along the $x$ axis and $y$ axis are compared by varying the numerical aperture (NA) and the position of substrate, respectively. Despite the fact that the resolution would be improved with a higher NA lens, Fig.~\ref{fig:Elliticity}(a) shows that with the silicon substrate \SI{1}{\micro\metre} below the focal plane (the dipole is placed at the focal point), the resolutions along $\hat{x}$ and $\hat{y}$ only have small differences. Then, fixing NA$=1$, Fig.~\ref{fig:Elliticity}(b) shows that the resolution has variations with different placements of the substrate. However, the maximum difference of FWHM w.r.t. the two observation directions is $0.048\lambda$, which is quite small. Therefore, the inference regarding resolution along x-direction can be easily generalized to the resolution along y-direction and only the lateral resolution along the $x$ direction is analyzed in the paper.  

More properties about the dyadic Green's function, first, the resolution would be the same if the polarization of two dipoles is exchanged. Second, the $x$-polarized dipole and the $y$-polarized dipole behaves similarly and have approximately the same resolution. This conclusion is made by realizing that when
observing field at $\vec{r}_\mathrm{ccd}$ for the $x$-polarized dipole and at $\vec{r}_\mathrm{ccd}^*$ for the $y$-polarized dipole, since $I_{xx}^{(1)}(\vec{r}_\mathrm{ccd})=I_{xx}^{(1)}(\vec{r}_\mathrm{ccd}^*)$, $I_{xx}^{(2)}(\vec{r}_\mathrm{ccd})=-I_{xx}^{(2)}(\vec{r}_\mathrm{ccd}^*)$, $I_{xy}(\vec{r}_\mathrm{ccd})=I_{xy}(\vec{r}_\mathrm{ccd}^*)$ and $I_{zx}(\vec{r}_\mathrm{ccd})=I_{zy}(\vec{r}_\mathrm{ccd}^*)$, we have the identity $I(\vec{r}_\mathrm{ccd})=I(\vec{r}_\mathrm{ccd}^*)$, i.e., the image of the $y$-polarized dipole would the same with the ${x}$-polarized dipole after exchanging the $x$ and $y$ coordinate of observation points. Together with the small ellipticity, we conclude that the lateral resolutions w.r.t. these two polarizations are with little difference. 

Thus, the following $4$ representative cases are studied, which are with polarizations
$\{\vec{p}_1=\hat{x}, \vec{p}_2=\hat{x}\}$, $\{\vec{p}_1=\hat{x}, \vec{p}_2=\hat{y}\}$, $\{\vec{p}_1=\hat{z}, \vec{p}_2=\hat{z}\}$, or 
$\{\vec{p}_1=\hat{x}, \vec{p}_2=\hat{z}\}$, $\vec{p}_i$ being the direction of the $i$-th dipole, $i=1,2$. The second case is due to the fact that it may have significantly higher resolution than the first case, as shown by Fig.~7.

\begin{figure}[!ht]
	\centering 
	\includegraphics[width=0.5\linewidth]{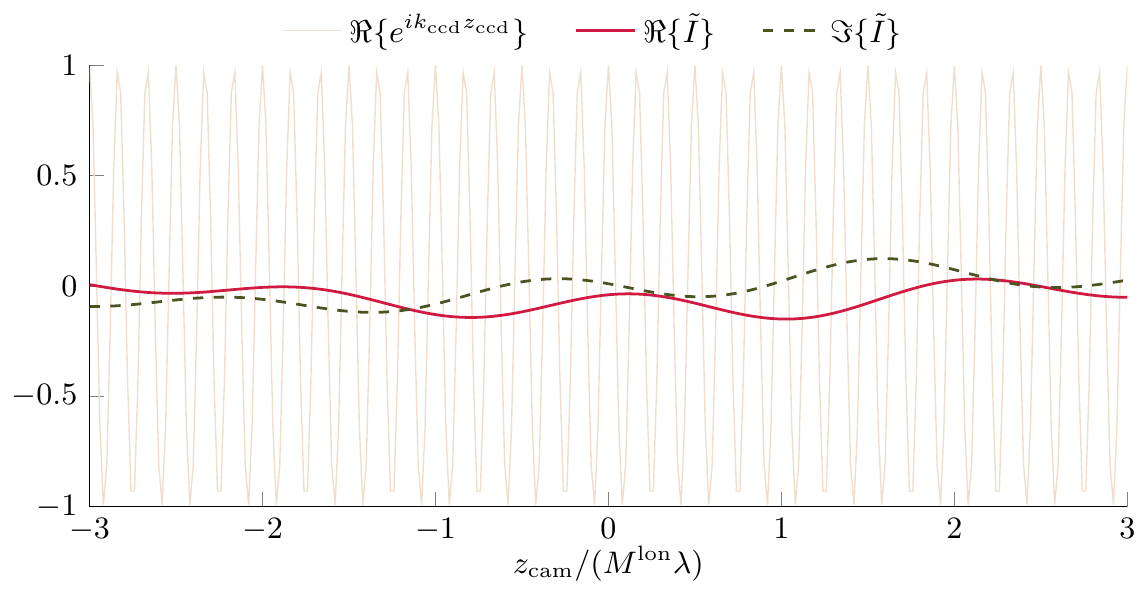}
	\caption{The combined field in Fig.~10(g) is a sinusoid with modulated amplitude.}
	\label{fig:sinePropery}
\end{figure}
{\bf Sinusoid-like observations along the optical axis:} The fields in Fig.~10(g) behave as a modulated sinusoid with a high (spatial) frequency. $z_\mathrm{cam}$ impacts the field solution through the exponential term in Eq.~(11). With the knowledge that $\cos\theta_\mathrm{cam}\approx1$ (due to the condition $f_\mathrm{cam}\gg f_\mathrm{obj}$) and $e^{ik_\mathrm{cam}z_\mathrm{cam}}$ is a sinusoid about $z_\mathrm{cam}$, it is straightforward to do the investigation with the rewritten form of Eq.(11),
\begin{equation}
\label{eq:combineIxx1_v2}
I_{xx}^{(1)} =e^{ik_\mathrm{cam}z_\mathrm{cam}}\int_{0}^{\theta_{\text{obj}}^{\max}}A(\theta_{\text{obj}})e^{ik_\mathrm{cam}(\cos\theta_\mathrm{cam}-1)z_\mathrm{cam}}\cos\bracket{k_\mathrm{obj}\cos\theta_{\text{obj}}\Delta z_\mathrm{obj}/2}d\theta_{\text{obj}}.
\end{equation}
Denote the integral in \eqref{eq:combineIxx1_v2} as $\tilde{I}$. Fig.~\ref{fig:sinePropery} shows the value of $\tilde{I}$, which also behaves as a sinusoid but with small amplitude and low frequency. Therefore, the combined field is a sinusoid whose amplitude is modulated by $\tilde{I}$.
\end{document}